\documentclass[useAMS,usenatbib,usegraphicx,pdftex]{mn2e}
\usepackage{graphicx}
\usepackage{amsmath,amssymb}
\usepackage{pslatex}
\usepackage{float}
\bibpunct{(}{)}{;}{a}{}{,}

\title[A variable ULX in M51a]{A variable ULX and possible IMBH candidate in M51a}
\author[Hannah M. Earnshaw, et al.]{\parbox{\textwidth}{Hannah M. Earnshaw$^{1}$\thanks{E-mail:
hannah.earnshaw@durham.ac.uk}, Timothy P. Roberts$^{1}$, Lucy M. Heil$^{2}$, Mar Mezcua$^{3,4}$, Dominic J. Walton$^{5}$, Chris Done${^1}$, Fiona A. Harrison$^{5}$, George B. Lansbury$^{1}$, Matthew J. Middleton$^{6}$, Andrew D. Sutton$^{7}$}\\
\\
\parbox{\textwidth}{
$^1$Centre for Extragalactic Astronomy, Department of Physics, Durham University, South Road, Durham, DH1 3LE, UK\\
$^2$Anton Pannekoek Institute, Science Park 904,1098 XH, Amsterdam, Netherlands\\
$^3$University of Montr\'{e}al, Pavillon Roger-Gaudry (D-428), 2900 boul. \'{E}douard-Montpetit, Montr\'{e}al QC, H3T 1J4, Canada\\
$^4$Harvard-Smithsonian Center for Astrophysics (CfA), 60 Garden Street, Cambridge, MA 02138, USA\\
$^5$California Institute of Technology, Pasadena, CA 91125, USA\\
$^6$Institute of Astronomy, University of Cambridge, Madingley Road, Cambridge CB3 0HA, UK\\
$^7$Astrophysics Office, NASA Marshall Space Flight Center, ZP12, Huntsville, AL 35812, USA
}}

\begin{document}

\date{}

\pagerange{\pageref{firstpage}--\pageref{lastpage}} \pubyear{}

\maketitle

\label{firstpage}

\begin{abstract}

ULX-7, in the northern spiral arm of M51, demonstrates unusual behaviour for an ultraluminous X-ray source, with a hard X-ray spectrum but very high short-term variability. This suggests that it is not in a typical ultraluminous state. We analyse the source using archival data from {\it XMM-Newton}, {\it Chandra} and {\it NuSTAR}, and by examining optical and radio data from {\it HST} and {\it VLA}. Our X-ray spectral analysis shows that the source has a hard power-law spectral shape with a photon index $\Gamma\sim1.5$, which persists despite the source's X-ray luminosity varying by over an order of magnitude. The power spectrum of the source features a break at $6.5^{+0.5}_{-1.1}\times10^{-3}$~Hz, from a low-frequency spectral index of $\alpha_1=-0.1^{+0.5}_{-0.2}$ to a high-frequency spectral index of $\alpha_2=0.65^{+0.05}_{-0.14}$, making it analogous to the low-frequency break found in the power spectra of low/hard state black holes (BHs). We can take a lower frequency limit for a corresponding high-frequency break to calculate a BH mass upper limit of $1.6\times10^3$\,M$_{\odot}$. Using the X-ray/radio fundamental plane we calculate another upper limit to the BH mass of $3.5\times10^4$\,M$_{\odot}$ for a BH in the low/hard state. The hard spectrum, high rms variability and mass limits are consistent with ULX-7 being an intermediate-mass BH; however we cannot exclude other interpretations of this source's interesting behaviour, most notably a neutron star with an extreme accretion rate.

\end{abstract}

\begin{keywords}
accretion, accretion discs -- black hole physics -- galaxies: individual: M51 -- X-rays: binaries -- X-rays: individual: M51 ULX-7
\end{keywords}

\section{Introduction}
\label{sec:intro}

Ultraluminous X-ray sources (ULXs) are point sources that are located away from the centre of their host galaxies and have an X-ray luminosity $L_{\rm X}>10^{39}$\,erg\,s$^{-1}$, the Eddington luminosity of a typical stellar-mass black hole (BH) with $M_{\rm BH}\sim10$\,M$_{\odot}$ (\citealt{ozel10}; see \citealt{feng11} for a review of ULXs). Many of these sources can be explained as stellar-mass BHs accreting at or close to the Eddington limit (e.g. \citealt{middleton13}). However for sources with $L_{\rm X}>3\times10^{39}$\,erg\,s$^{-1}$, this is often not sufficient. Given that their non-nuclear nature rules out the sources being active galactic nuclei (AGNs), and assuming that they are not background sources erroneously associated with the galaxy, they require one of two alternative explanations. Either they are BHs of unusually high mass -- that is, intermediate-mass BHs (IMBHs) with $10^2 \lesssim M_{\rm BH} \lesssim 10^5$\,M$_{\odot}$ \citep{colbert99} -- or they exhibit an extreme accretion mechanism beyond the standard thin disc scenario, such as super-Eddington accretion \citep{poutanen07} and/or geometrically beamed emission \citep{king01}.

The origin of IMBHs requires exotic formation scenarios, such as the collapse of early-universe population III stars \citep{madau01}. This, and the results of studies implying that all but the brightest sources of the ULX population make up the high luminosity tail of the high mass X-ray binary (HMXB) luminosity function in spiral galaxies and the low mass X-ray binary (LMXB) luminosity function in elliptical galaxies \citep{swartz04, swartz11, walton11,mineo12}, points to the majority of ULXs being super-Eddington accreting stellar-mass sources. This is strengthened by results from the spectral analysis of these sources, which exhibit a characteristic two-component spectrum with a soft excess and a high energy downturn at $\sim3-5$\,keV that is not found in sub-Eddington accretion states (e.g. \citealt{stobbart06, gladstone09, miyawaki09}). This downturn has been confirmed by observations of ULXs using {\it NuSTAR} (e.g. \citealt{bachetti13, walton15}).

There are currently a small number of good candidates for IMBHs. These tend to be objects too luminous to be explained by super-Eddington accretion onto stellar-mass BHs, in particular hyper-luminous sources (HLXs; $L_{\rm X}>10^{41}$\,erg\,s$^{-1}$; \citealt{gao03, farrell09, sutton12}), or ULXs with powerful radio jets (e.g. \citealt{mezcua13, mezcua15}). For example one of the best HLX candidates, ESO 243-49 HLX-1 (henceforth HLX-1), has been observed in different spectral states similar to the hard and thermal dominated states seen in stellar-mass BH binaries (BHBs; \citealt{godet09, servillat11}). This strongly supports the interpretation of HLX-1 as a sub-Eddington accreting source scaled up to higher masses. If IMBHs do exist, then we might expect that some are accreting at lower rates and have a similar luminosity to stellar-mass ULXs, although distinguishing them from stellar-mass ULXs would be difficult and dependent on the spectral and timing properties of the source.

In order to study the properties of ULXs as a population, and to find the most likely candidates for IMBHs, we created a new, clean catalogue of candidate ULXs (\citeauthor{earnshaw16} in prep.). This matched the 3XMM-DR4 data release of the {\it XMM-Newton} Serendipitous Sky Survey \citep{rosen15} with the Third Reference Catalog of Bright Galaxies \citep{devaucouleurs91}, using a method similar to \citet{walton11}, with a number of improvements to reduce contamination by camera artefacts and background sources. This catalogue contains 331 candidate ULXs, which we searched for sources of interest on the basis of luminosity or variability. We found a small number of highly variable ULXs, including one particularly interesting source in M51.

The interacting galaxy system M51 (NGC~5194/5, also known as the Whirlpool Galaxy) is a pair of galaxies at a distance of 7.85\,Mpc\footnote{The mean distance given by the NASA/IPAC Extragalactic Database (http://ned.ipac.caltech.edu/).}, containing the face-on spiral galaxy M51a which has high rates of star formation and a large population of X-ray sources, including nine ULXs \citep{terashima06}. One such ULX located in the northern spiral arm, henceforth referred to as ULX-7, has a hard X-ray spectrum and high levels of variability. Its variability was first investigated by \citet{liu2002}, who found a tentative period of 7620\,s. Later studies (e.g \citealt{dewangan05, terashima04, terashima06}) found significant long- and short-term variability, although they did not find a period, suggesting that the variability is instead due to aperiodic noise from stochastic processes. The source is near to a young massive star cluster with age $T\sim12$\,Myr \citep{abolmasov07} and has previously been found to have a changing spectral shape by \citet{yoshida10}, ranging from fairly flat ($\Gamma<1.5$) to soft ($\Gamma\sim2-3$), although any contribution from the host galaxy to the emission was not considered in their study.

While they often demonstrate spectral variability between observations (e.g. \citealt{kajava09}), strong short term variability is not a common feature of ULXs (e.g. \citealt{feng05, heil09}). In the broadened disc/hard ultraluminous/soft ultraluminous classification of ULX accretion regimes, high ($>10\%$) fractional variability is predominantly seen in the soft ultraluminous state in which the X-ray spectrum is dominated by soft thermal emission \citep{sutton13}. Furthermore, the variability is limited to the hard component of emission \citep{middleton15}. A proposed mechanism for this is a clumpy wind that would be expected to be driven away from the disc by intense radiation pressure in super-Eddington accretion scenarios, and to intermittently obscure the hard central source in high-inclination systems, causing the spectrum to be dominated by soft thermal emission and variability to be imprinted on the hard emission component (e.g. \citealt{middleton11, middleton15}). However in the case of ULX-7, the spectrum is hard, which suggests that this source does not fit this model and may be in an accretion state more analogous to the low/hard state of stellar-mass BHs -- in which case, this object might instead be a candidate IMBH, albeit emitting at a lower luminosity than other candidate IMBHs we are aware of to date. 

Here we conduct our own analysis of ULX-7, examining its X-ray spectral and timing properties, as well as optical and radio data, to attempt to better characterise this fascinating source. In Section~\ref{sec:data} we detail the reduction of archival data from {\it XMM-Newton}, {\it Chandra}, {\it NuSTAR}, the Hubble Space Telescope ({\it HST}) and the Very Large Array ({\it VLA}) telescopes. We present the results of our analysis in Section~\ref{sec:results}, and discuss the possibility of this source being a background AGN, a neutron star or an IMBH in Section~\ref{sec:disc}, before presenting our conclusions in Section~\ref{sec:conc}.

\section{Reduction of Archival Data}
\label{sec:data}

In this paper we will investigate this object from a multi-wavelength perspective, using archival data from a range of missions. We assume a source position of 13:30:01.0~+47:13:44 (J2000; \citealt{kilgard05}).

\subsection{X-ray Observations}
\label{sec:xray}

There were six observations of M51 by the {\it XMM-Newton} observatory over the course of eight years, between 2003 and 2011. The durations of observations with {\it XMM-Newton} are limited by visibility due to the position of M51 in the sky and the orbit of the telescope. The longest observation to date is $\sim$52\,ks. Data reduction was performed using {\sc v13.5.0} of the {\it XMM-Newton} Scientific Analysis System (SAS) and up-to-date calibration files. We used {\sc epproc} and {\sc emproc} to produce calibrated event lists for the pn and MOS detectors. The event lists were filtered for high-energy background flaring in accordance with the standard {\it XMM-Newton} SAS threads\footnote{See the SAS User Manual at http://xmm.esac.esa.int/sas}, excluding intervals for which the $>10$\,keV count rate was $\geq0.35$\,ct\,s$^{-1}$ in the EPIC-MOS data and the 10--12\,keV count rate was $\geq0.4$\,ct\,s$^{-1}$ in the EPIC-pn data. 

{\it XMM-Newton} spectra and light curves were extracted using {\sc evselect} from 20\,arcsecond radius regions around the source, filtering for {\tt pattern}$\leqslant12$ for the EPIC-MOS camera and {\tt pattern}$\le4$ for the EPIC-pn camera. Background counts were extracted from an equally-sized region outside of the galaxy on the same chip, at a similar distance from the readout node. Redistribution matrices and auxilary response files were generated using {\sc rmfgen} and {\sc arfgen} respectively, and spectral data were grouped into bins of at least 25 counts, making sure not to oversample {\it XMM-Newton}'s intrinsic energy resolution by a factor more than three. Corrected lightcurves for EPIC-MOS and EPIC-pn were generated using {\sc epiclccorr} with a bin size of 50\,s and added together, using the same start and end times. All six observations have a quality warning flag due to the source being located within bright extended emission, which we characterise in Section~\ref{sec:xray}.

While {\it Chandra} does not have the collecting area of {\it XMM-Newton}, it is not limited by visibility in the same way, and its orbit allows for far longer observations of M51. There have been eleven {\it Chandra} observations of M51 taken over the course of twelve years, between 2000 and 2012, the longest being $\sim$190\,ks (see Table~\ref{tab:obs}) during a set of five deep observations in 2012. The {\it Chandra} data were reduced using {\sc v4.7} of the Chandra Interactive Analysis of Observations (CIAO) software package and reprocessed to produce up-to-date event lists.

{\it Chandra} spectra and light curves were extracted using the {\sc specextract} and {\sc dmextract} routines respectively from 3\,arcsecond radius regions, with the same binning as the {\it XMM-Newton} data. Since ULX-7 is a point source, we set {\tt weight=no} and {\tt correctpsf=yes}. Background counts were collected from an annulus around the source between 3 and 20\,arcseconds. This same annulus was also used to characterise diffuse emission surrounding the object in order to correct the {\it XMM-Newton} spectra -- the background region in this case was taken from an equally-sized region outside of the galaxy. Between {\it XMM-Newton} and {\it Chandra}, we can examine the long-term variability of the source, as well as its short-term properties.

The advent of the {\it NuSTAR} mission allows us to probe spectral energies of $>10$\,keV for resolved sources. To date there is one observation of M51 using {\it NuSTAR}, performed in 2012, in which ULX-7 is detected along with the low-luminosity AGN and one other ULX in the galaxy. We reduced the \textit{NuSTAR} data using the standard pipeline, \textsc{nupipeline}, part of the \textit{NuSTAR} Data Analysis Software (\textsc{NuSTARDAS}, v1.4.1; included in the \textsc{HEASOFT} distribution), with the instrumental calibration files from {\sc caldb v}20140414. The unfiltered event files were cleaned with the standard depth correction, significantly reducing the internal high-energy background, and passages through the South Atlantic Anomaly were removed. Source spectra and instrumental responses were produced for each of the two focal plane modules (FPMA/B) using \textsc{nuproducts}. Source spectra were extracted from a circular region of radius 25\,arcseconds in order to avoid contamination from a potential nearby X-ray source, while background was estimated from a much larger region on the same detector as the source, avoiding all the other bright X-ray sources in M51. In order to maximise the good exposure, in addition to the standard (mode 1) data, we also reduce the available mode 6 data; see Walton et al. and Fuerst et al. (in preparation) for a description of \textit{NuSTAR} mode 6. This provides an additional $\sim$15\% exposure, resulting in a total on-source time of 19\,ks per FPM. Finally, owing to the low signal-to-noise, we combined the data from FPMA and FPMB using \textsc{addascaspec}. The resulting \textit{NuSTAR} spectrum provides a detection up to $\sim$20--25\,keV, and is rebinned to a minimum of 20 counts per bin for our spectral analysis.

A list of all X-ray observations is presented in Table~\ref{tab:obs}. We calculate the flux between 0.3--10\,keV (3.0--10\,keV and 3.0--20\,keV for the {\it NuSTAR} data) using the best-fitting absorbed power-law model if there are sufficient counts for a spectral fit. In the cases where there are a small number of data points we use WebPIMMS\footnote{https://heasarc.gsfc.nasa.gov/cgi-bin/Tools/w3pimms/w3pimms.pl} and the count rate to predict the flux, assuming a photon index of $\Gamma=1.5$ (the average photon index of the best-fitting models to the data we were able to fit -- see Section~\ref{sec:spec}). In the case of {\it XMM-Newton} observation X6, it should be noted that the flux is likely dominated by the surrounding diffuse emission rather than the source itself. The source flux varies between $3.0\times10^{-14}$\,erg\,cm$^{-2}$\,s$^{-1}$ and $8.6\times10^{-13}$\,erg\,cm$^{-2}$\,s$^{-1}$ in the 0.3--10\,keV energy band over the course of all observations (i.e. its luminosity varies between $2.2\times10^{38}$\,erg\,s$^{-1}$ and $5.1\times10^{39}$\,erg\,s$^{-1}$). This is an unusually high amount of flux variation for a ULX, even if we disregard fluxes calculated using WebPIMMS, in which case we still see variation of over an order of magnitude.

\begin{table*}
\begin{minipage}{120mm}
\caption{Dates, durations and fluxes for X-ray observations of M51 with {\it XMM-Newton}, {\it Chandra} and {\it NuSTAR}.} \label{tab:obs}
\begin{center}
\begin{tabular}{cccccc}
  \hline
  ID$^a$ & Observation ID & Instrument & Observation Date & Exposure$^b$ & Flux$^c$ \\
   &  &  &  & (ks) & $\;\;$($\times 10^{-13}$\,erg\,cm$^{-2}$\,s$^{-1}$) \\
  \hline
  \multicolumn{6}{c}{\textit{\textbf{XMM-Newton}}} \\
  \rule{0pt}{2.5ex}
  X1 & 0112840201 & MOS1 & 2003-01-15 & 20.66 & $1.9 \pm 0.1$ \\
    & & MOS2 &  & 20.67 &  \\
    & & pn &  & 19.05 &  \\
  X2 & 0212480801 & MOS1 & 2005-07-01 & 35.04 & $7.6 \pm 0.2$ \\
    & & MOS2 &  & 35.77 &  \\
    & & pn &  & 24.94 &  \\
  X3 & 0303420101 & MOS1 & 2006-05-20 & 39.60 & $5.9 \pm 0.1$ \\
    & & MOS2 &  & 39.66 &  \\
    & & pn &  & 30.90 &  \\
  X4 & 0303420201 & MOS1 & 2006-05-24 & 29.77 & $8.5 \pm 0.2$ \\
    & & MOS2 &  & 29.76 &  \\
    & & pn &  & 23.18 &  \\
  X5 & 0677980701 & MOS1 & 2011-06-07 & 9.80 & $5.2 \pm 0.4$ \\
    & & MOS2 &  & 9.51 &  \\
    & & pn &  & 5.13 &  \\
  X6 & 0677980801 & MOS1 & 2011-06-11 & 1.60 & $1.1 \pm 0.1^d$ \\
    & & MOS2 &  & 1.60 &  \\
    & & pn &  & 2.30 &  \\
  \multicolumn{6}{c}{\textit{\textbf{Chandra}}} \\
  \rule{0pt}{2.5ex}
  C1 & 354 & ACIS-S & 2000-06-20 & 15.05 & $4.5 \pm 0.5$ \\
  C2 & 1622 & ACIS-S & 2001-06-23 & 27.15 & $0.5 \pm 0.1^d$ \\
  C3 & 3932 & ACIS-S & 2003-08-08 & 48.61 & $4.7 \pm 0.2$ \\
  C4 & 12562 & ACIS-S & 2011-06-12 & 9.75 & $0.8 \pm 0.1^d$ \\
  C5 & 12668 & ACIS-S & 2011-07-03 & 10.12 & $4.0\pm0.6$ \\
  C6 & 13812 & ACIS-S & 2012-09-12 & 159.54 & $7.1 \pm 0.2$ \\
  C7 & 13813 & ACIS-S & 2012-09-09 & 181.57 & $8.6 \pm 0.2$ \\
  C8 & 13814 & ACIS-S & 2012-09-22 & 192.36 & $5.5 \pm 0.1$ \\
  C9 & 13815 & ACIS-S & 2012-09-23 & 68.07 & $3.6 \pm 0.2$ \\
  C10 & 13816 & ACIS-S & 2012-09-26 & 74.07 & $0.8 \pm 0.1$ \\
  C11 & 15496 & ACIS-S & 2012-09-19 & 41.51 & $6.0 \pm 0.3$ \\
  C12 & 15553 & ACIS-S & 2012-10-10 & 38.07 & $0.3 \pm 0.1^d$ \\
  \multicolumn{6}{c}{\textit{\textbf{NuSTAR}}} \\
  \rule{0pt}{2.5ex}
  N1 & 60002038002 & FPMA/B & 2012-10-29 & 19 & $3.8 \pm 0.7$ \\
   & & & & & $5.6 \pm 0.9$ \\
  \hline
\end{tabular}
\end{center}
$^a$A short ID used elsewhere within this paper for clarity. \\
$^b$Sum of the good time intervals after removal of background flaring events. \\
$^c$Deabsorbed flux in the energy range 0.3--10\,keV for {\it XMM-Newton} and {\it Chandra}, and the ranges 3--10\,keV and 3--20\,keV for {\it NuSTAR}. The flux is determined from the best-fitting power-law model, excluding contribution from soft diffuse emission in the {\it XMM-Newton} observations (see Section~\ref{sec:spec}). The {\it NuSTAR} fluxes are calculated from the best-fitting power-law model (with $N_{\rm H}=1\times10^{21}$\,cm$^{-2}$) to the 3--10\,keV data and the 3--20\,keV data respectively. \\
$^d$Where there is insufficient data to calculate flux from a best-fitting model, we use WebPIMMS to find the deabsorbed flux, using the detected count rate and a power-law model with $N_{\rm H} = 1\times10^{21}$\,cm$^{-2}$ and $\Gamma=1.5$.
\end{minipage}
\end{table*}

\subsection{Radio Observations}
\label{sec:radio}

\begin{figure}
\begin{center}
\includegraphics[width=8cm]{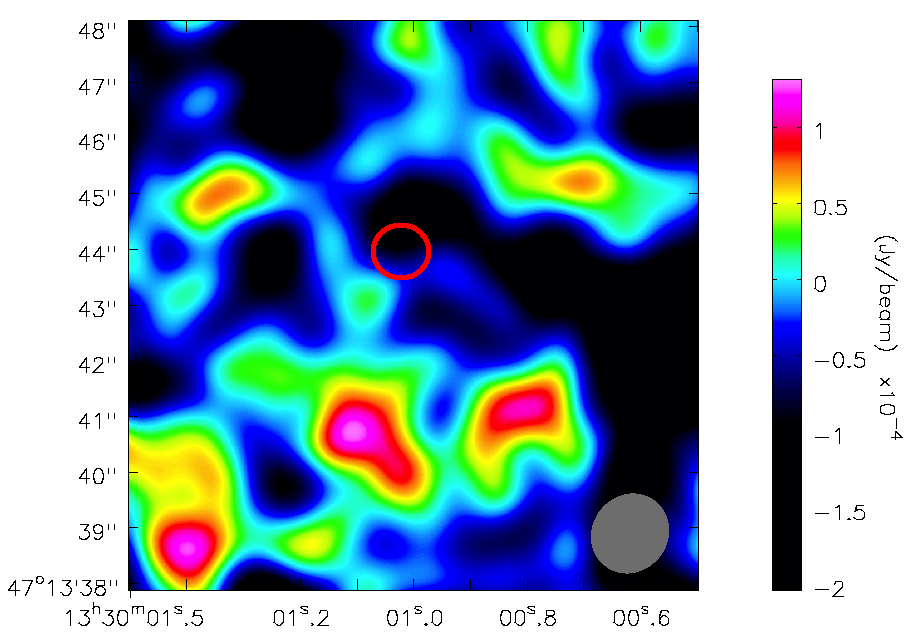}
\end{center}
\caption{The 1.5~GHz radio image surrounding the source location, which is marked with a 1~arcsecond error circle (red) and does not coincide with a radio detection. The rms is $8.73\times10^{-5}$\,Jy/beam, therefore the fluctuations shown are all consistent with noise. The beam size is marked in grey.}
\label{fig:radio}
\end{figure}

In order to search for core radio emission from the ULX, we retrieved archival {\it VLA} A-array data at 1.5\,GHz (project 11A­142, August 2011). The data flagging and calibration was performed following standard procedures with the Common Astronomy Software Applications ({\sc CASA}) software. The data were calibrated in amplitude using 3C286 as flux calibrator, while delay and phase solutions were derived from the phase calibrator J1327+4326 and interpolated and applied to the target source. The calibrated data were imaged in {\sc CASA} using the Cotton­Schwab algorithm and natural weighting. The resulting beam has a size of 1.5\,arcseconds\,$\times$\,1.4\,arcseconds oriented at a position angle of ­36.9\,deg. No radio emission is detected at the \textit{Chandra} position of the source within a positional error of 1\,arcsecond. An upper limit on the 1.5\,GHz radio flux density of 87\,$\mu$Jy\,beam$^{-­1}$ is derived from the local rms at the \textit{Chandra} position. The 1.5\,GHz radio image is shown in Fig.~\ref{fig:radio}. Other studies of the radio emission in M51 have also not detected a counterpart to ULX-7 \citep{maddox07, rampadarath15}.

\subsection{Optical Observations}
\label{sec:optical}

M51 has been well observed by {\it HST} over the course of the mission's lifetime, and was mapped in 2005 with the ACS/WFC camera as part of the Hubble Heritage project. We collected pre-processed data from the Hubble Legacy Archive, made up of exposures combined using the {\sc MultiDrizzle} routine. We used images in the F435W ({\it B}), F555W ({\it V}) and F814W ({\it I}) bands to locate possible optical counterparts to ULX-7. The 90\% confidence circle for the {\it Chandra} ACIS-S instrument is 0.6\,arcseconds, however it is also necessary to align the relative astrometry of the {\it HST} and {\it Chandra} images. We did this by selecting {\it 2MASS} objects within the M51 field and using the {\sc iraf} tools {\sc ccfind}, {\sc ccmap} and {\sc ccsetwcs} to find the necessary corrections to the right ascension and declination. We found an offset of 0.1\,arcseconds in the right ascension direction and 0.7\,arcseconds in declination.

The source is located near to a young star cluster and has a number of possible optical counterparts. We performed photometry on these objects using the {\sc DAOPHOT II/ALLSTAR} software package \citep{daophot}, a PSF-fitting routine (see Section~\ref{sec:hubble}), although due to the crowded nature of the field, we were only able to obtain limited constraints on the magnitudes in each band. Where the magnitude of an object was unconstrained, we used the various sources of detector noise to place a lower limit on the magnitude.

\vspace{-1cm}
\section{Analysis \& Results}
\label{sec:results}

\begin{figure*}
\begin{center}
\includegraphics[height=10cm]{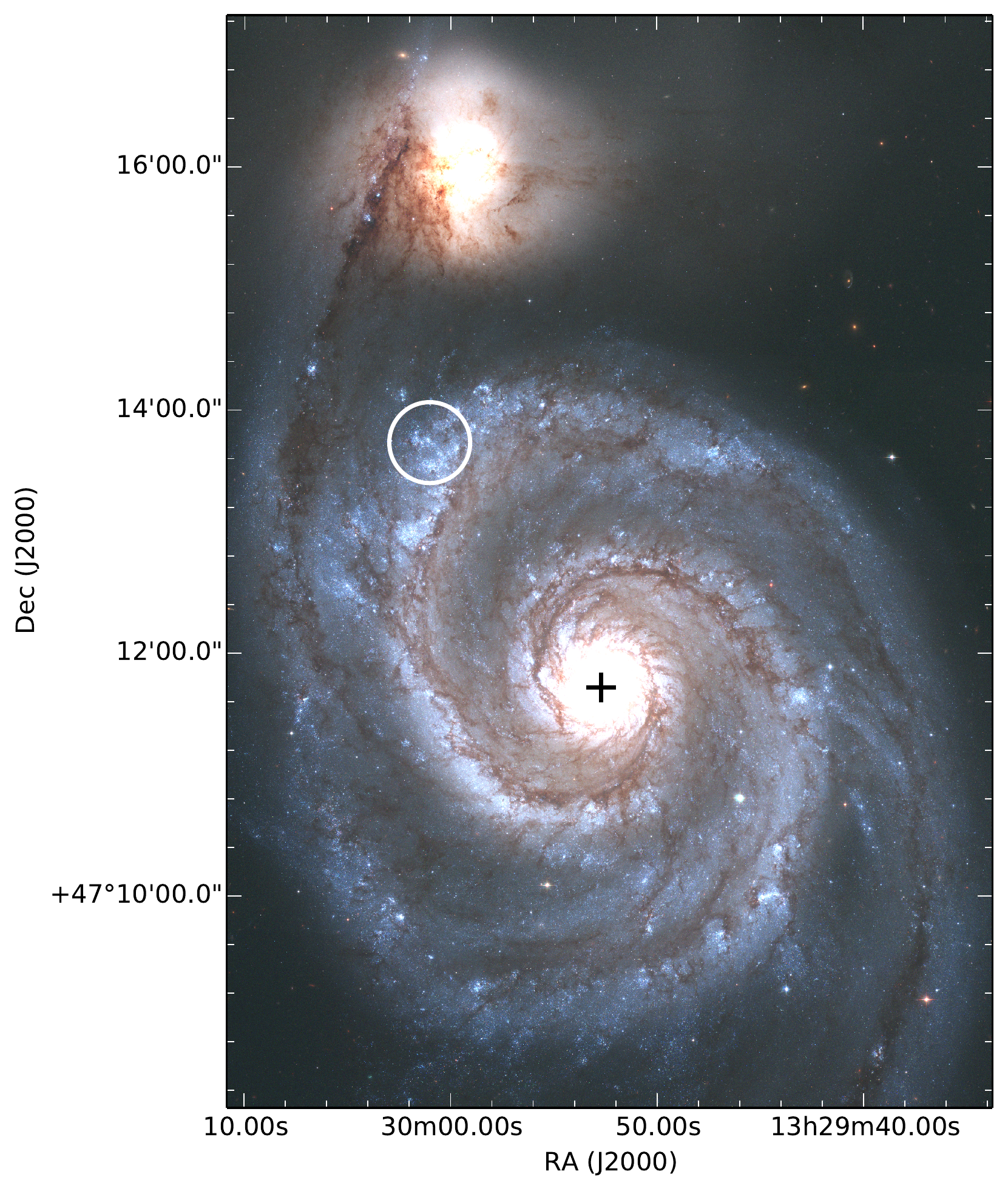}
\hspace{0.1cm}
\includegraphics[height=10cm]{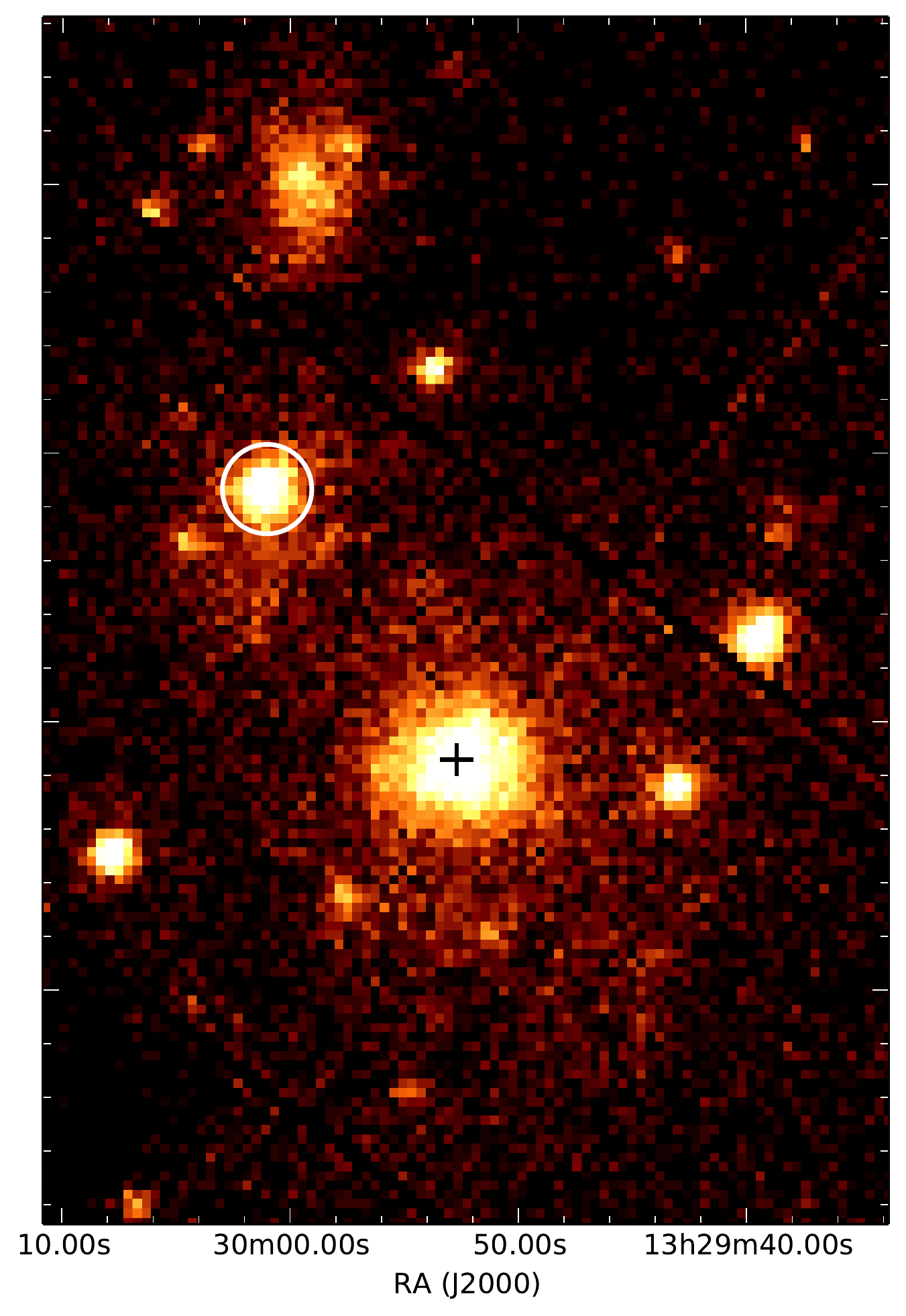}
\hspace{0.1cm}
\includegraphics[height=10cm]{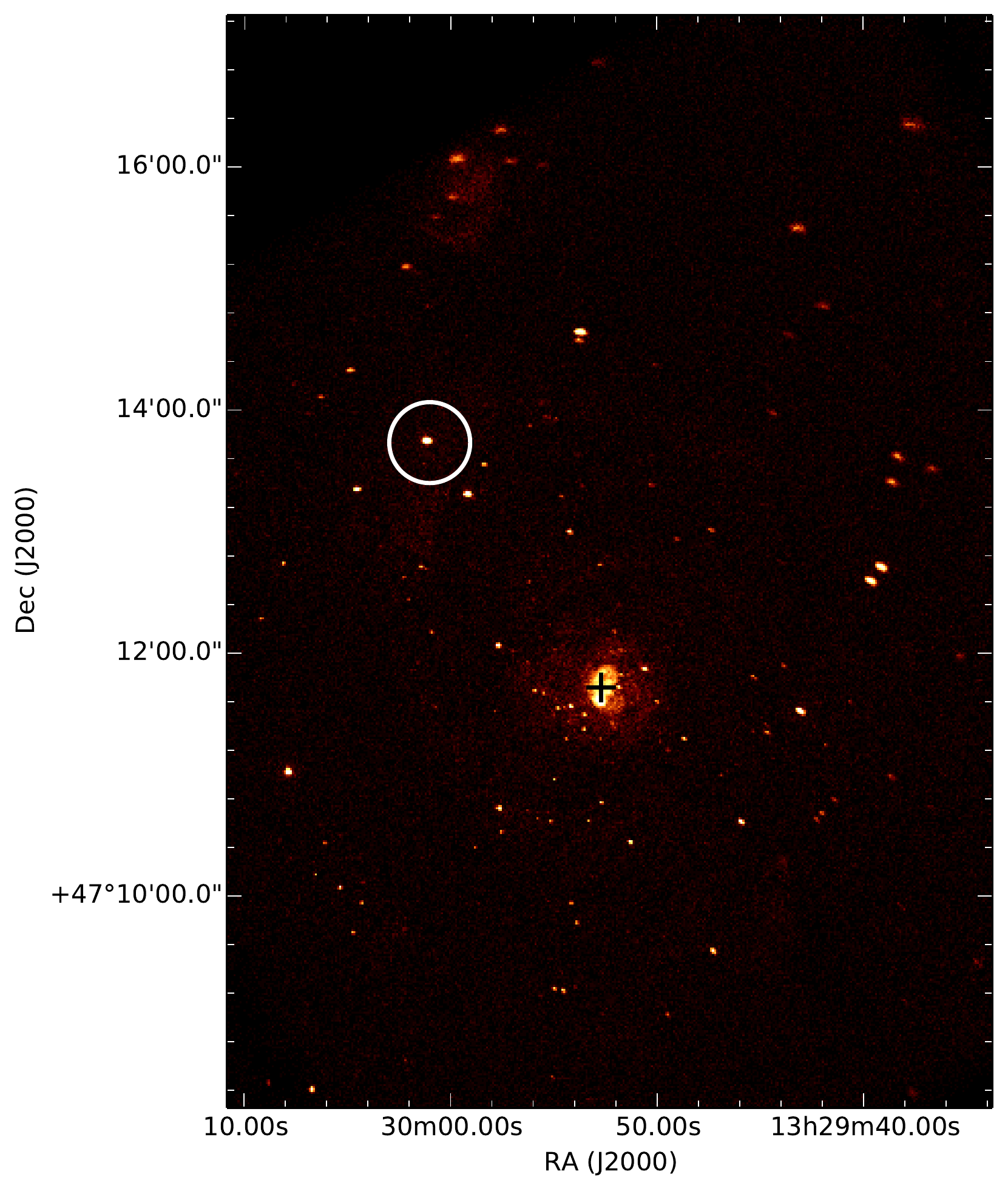}
\hspace{0.1cm}
\includegraphics[height=10cm]{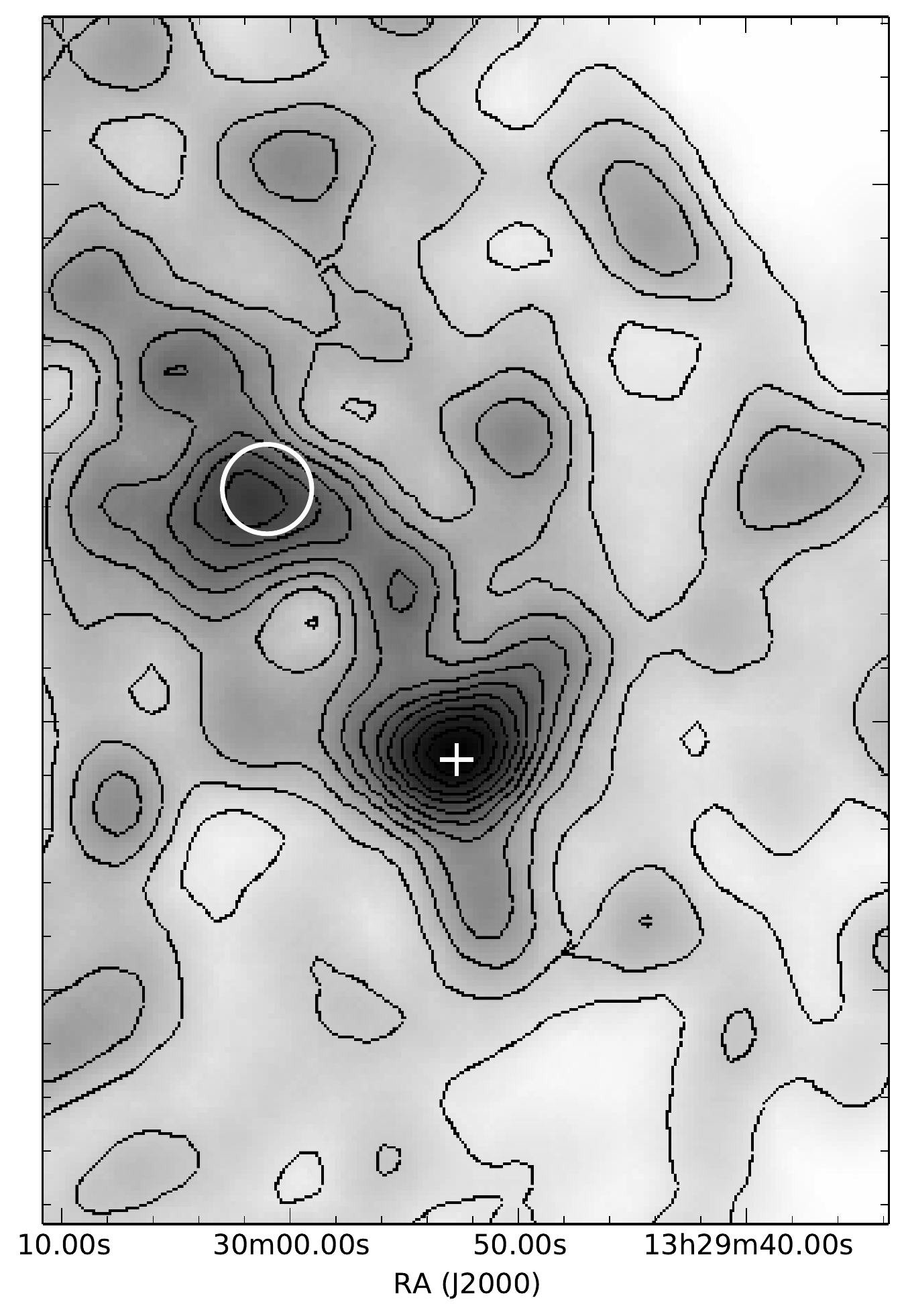}
\end{center}
\caption{Images of the M51 system, centred for convenience on 13:29:52.3~+47:12:45.3 (J2000). In all images, the position of the centre of M51a is marked with a cross, and ULX-7 is indicated by a 20\,arcsecond radius white circle. {\it Top left}, {\it HST} true-colour image with the red, green and blue channels corresponding to the F814W, F555W and F435W bands respectively. {\it Top right}, {\it XMM-Newton} EPIC-pn image in the energy range 0.3--10\,keV from observation X4. {\it Bottom left}, {\it Chandra} ACIS-S image in the energy range 0.3--10\,keV from observation C8. {\it Bottom right}, {\it NuSTAR} image in the energy range 3--24\,keV, smoothed with a 14\,arcsecond Gaussian and with contours to aid visibility only.}
\label{fig:image}
\end{figure*}

We analysed the archival data described in Section~\ref{sec:data} in order to determine the properties of ULX-7. Optical and X-ray images of M51 from the {\it HST}, {\it XMM-Newton}, {\it Chandra} and {\it NuSTAR} telescopes, along with the location of ULX-7, are shown in Fig.~\ref{fig:image}. The source lies within diffuse X-ray emission in the northern spiral arm of its host galaxy.

\subsection{X-Ray Imaging \& Spectral Analysis}
\label{sec:spec}

\begin{table*}
\begin{minipage}{116mm}
\caption{The temperature, flux and $\chi^2$ goodness of fit for the two {\tt mekal} components used to fit the diffuse emission around ULX-7 in the deepest {\it Chandra} observations.} \label{tab:diff}
\begin{center}
\begin{tabular}{cccccc}
  \hline
  ID$^a$ & $kT_1$ & $F_1$ & $kT_2$ & $F_2$ & $\chi^2$/dof \\
   & (keV) & ($\times 10^{-14}$ erg\,cm$^{-2}$\,s$^{-1}$) & (keV) & ($\times 10^{-14}$ erg\,cm$^{-2}$\,s$^{-1}$) & \\
  \hline
  C6 & $0.25^{+0.08}_{-0.07}$ & $1.3 \pm 0.3$ &$0.6^{+0.4}_{-0.1}$ & $1.3 \pm 0.3$ & $54.3/60$ \\
  C7 & $0.22^{+0.05}_{-0.04}$ & $1.5 \pm 0.4$ & $0.7^{+0.2}_{-0.1}$ & $1.0 \pm 0.2$ & $84.2/73$ \\
  C8 & $0.17^{+0.10}_{-0.09}$ & $1.0 \pm 0.3$ & $0.4^{+0.2}_{-0.1}$ & $1.4 \pm 0.2$ & $76.2/71$ \\
  C9 & $<0.27$ & $2.5 \pm 0.4$ & $<1.0$ & $0.5 \pm 0.2$ & $14.1/21$ \\
  C10 & $0.24^{+0.05}_{-0.06}$ & $2.0 \pm 0.4$ & $0.8^{+0.6}_{-0.2}$ & $0.5 \pm 0.3$ & $34.4/24$ \\
  \rule{0pt}{4ex}
  All$^b$ & $0.26^{+0.03}_{-0.05}$ & $1.9 \pm 0.1$ & $0.8\pm0.2$ & $0.6 \pm 0.1$ & $279.9/258$ \\
  \hline
\end{tabular}
\end{center}
$^a$The short observation ID as defined in Table~\ref{tab:obs}.
$^b$The best-fitting parameters when fitting all five observations simultaneously.
\end{minipage}
\end{table*}

The {\it XMM-Newton} image of the source and its environment (see Fig.~\ref{fig:image}) shows that it lies within extended diffuse emission. Therefore the {\it XMM-Newton} source spectra are likely to be contaminated by a soft thermal component. In order to characterise this component, we first examine archival data from the {\it Chandra} observatory, since its high spatial resolving power allows us to separate out the spectra of the source and of the surrounding gas.

To obtain sufficient counts from the {\it Chandra} data for analysis, we used only the five observations with exposure time $>50$\,ks. We extracted diffuse emission spectra from an annulus with an inner radius of 3\,arcseconds around the source, and an outer radius of 20\,arcseconds to be the same as the {\it XMM-Newton} footprint used for source analysis. There are no resolved point sources within the annulus. We took a background spectrum from a 20\,arcsecond radius region centred to the north of the galaxy in an area with minimal diffuse emission. % background centred on 13:29:46 47:14:40

All spectral fitting was performed with {\sc v12} of {\sc XSPEC} \citep{xspec}, and all {\it Chandra} and {\it XMM-Newton} observations are fitted in the 0.3--10\,keV energy range with errors given at 90\% confidence intervals. The data is binned (see Section~\ref{sec:xray}) such that fitting can be performed using $\chi^2$ minimisation, and $\chi^2$ statistics used to determine the goodness-of-fit. The abundance tables of \citet{tbabs} are used throughout.

The diffuse emission spectra were well-fitted using two {\tt mekal} thermal plasma components: a cooler component at $\sim0.2$~keV and a second warmer component at $\sim0.7$~keV, consistent with previous studies into the diffuse emission of the galaxy (e.g. \citealt{owen09}). We also detected hard emission, requiring an additional hard component in the spectrum since attempting to fit the data without it causes one of the {\tt mekal} components to take on an unrealistically high temperature. This hard component may be due to unresolved hard sources within the annulus, therefore we fitted it with an absorbed power-law ({\tt tbabs*powerlaw}), allowing the photon index to vary. We set the hydrogen column density to $N_{\rm H}=1\times10^{21}$\,cm$^{-2}$ since preliminary fits to the ULX-7 source spectrum gave $N_{\rm H}$ of approximately this value and we would expect absorption by the surrounding interstellar medium (ISM) to be similar in the near vicinity. The power-law has $\Gamma\sim$1--2 and would contribute $<0.1\%$ of the total flux when combined with the source spectrum. For this reason, we expect that its effect on the spectrum of ULX-7 is negligible, so we do not include it in our characterisation of the diffuse emission itself.

\begin{figure}
\begin{center}
\includegraphics[width=9cm]{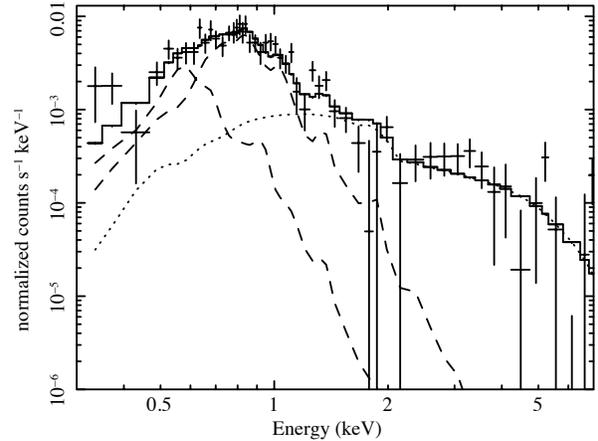}
\end{center}
\vspace{-0.2cm}
\caption{The X-ray spectrum of the diffuse emission surrounding ULX-7 from observation C8, together with the best-fitting {\tt mekal+mekal+tbabs*powerlaw} model, with {\tt mekal} parameters as given in Table~\ref{tab:diff}. The {\tt mekal} components are plotted with dashed lines and the {\tt tbabs*powerlaw} component with a dotted line. Events are grouped into 20 counts per bin.}
\label{fig:gas}
\end{figure}

The fit results for the diffuse emission are given in Table~\ref{tab:diff}. Given that the temperature parameters are all consistent within the errors, and that we do not expect the diffuse emission to vary between observations if it originates in the ISM of M51, we performed a simultaneous fit of all five observations and used the best-fitting parameters (see Table~\ref{tab:diff}) when fitting the {\it XMM-Newton} source spectra. An example of the diffuse emission spectrum is shown in Fig.~\ref{fig:gas}.

\begin{figure*}
\begin{center}
\includegraphics[width=9.3cm]{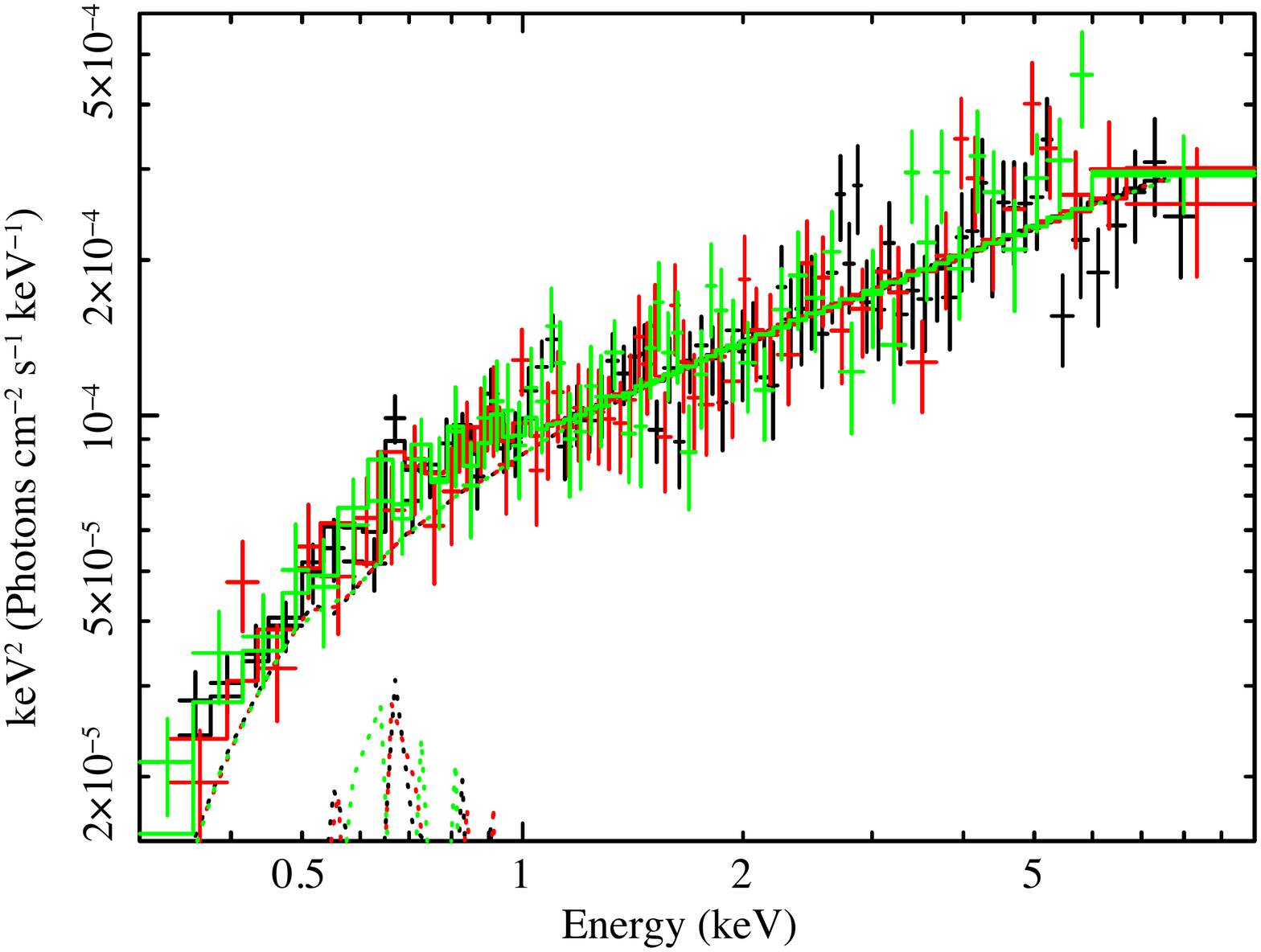}
\hspace{-1.2cm}
\includegraphics[width=9.3cm]{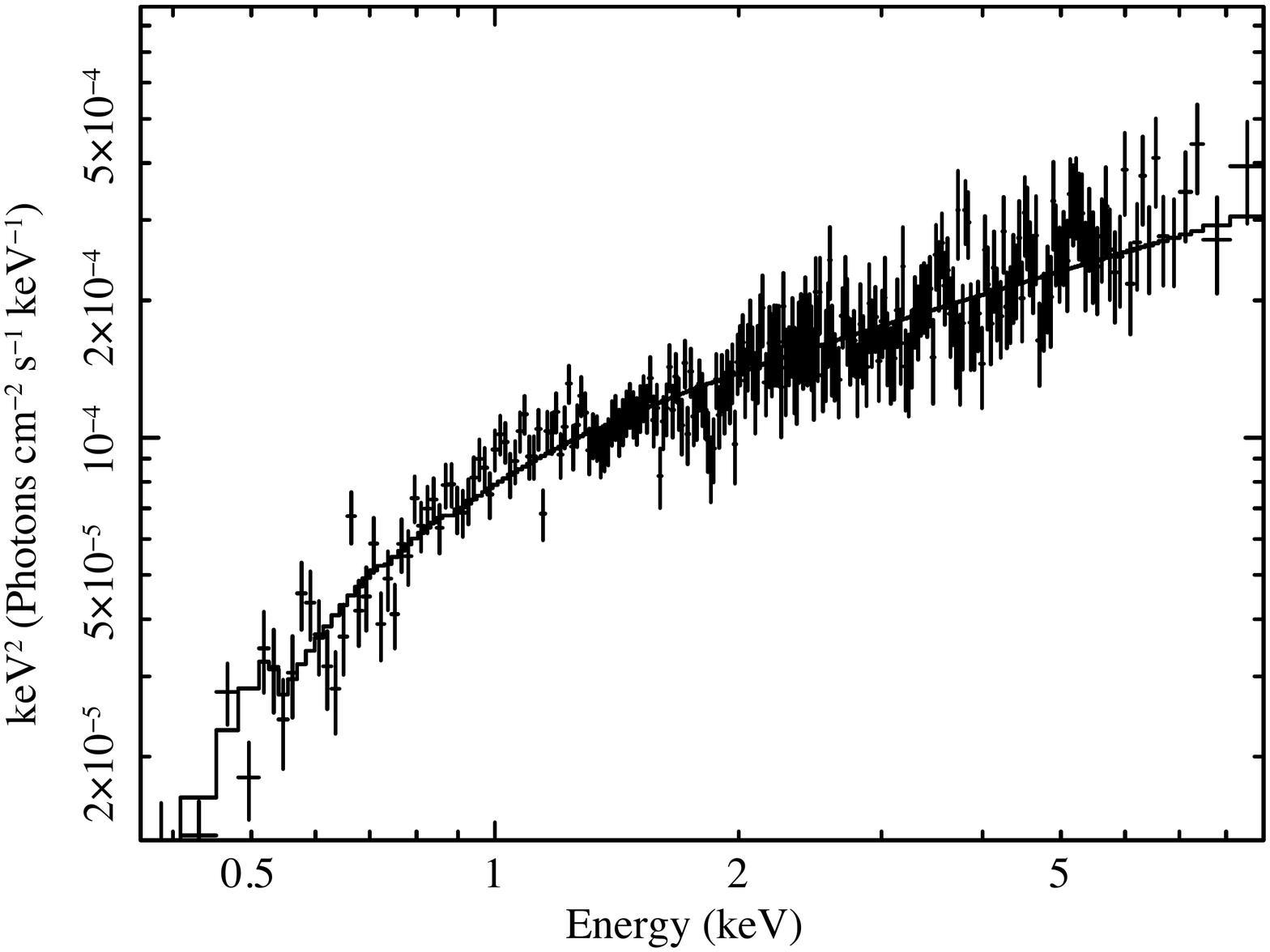}\\
\vspace{-0.8cm}
\includegraphics[width=9.3cm]{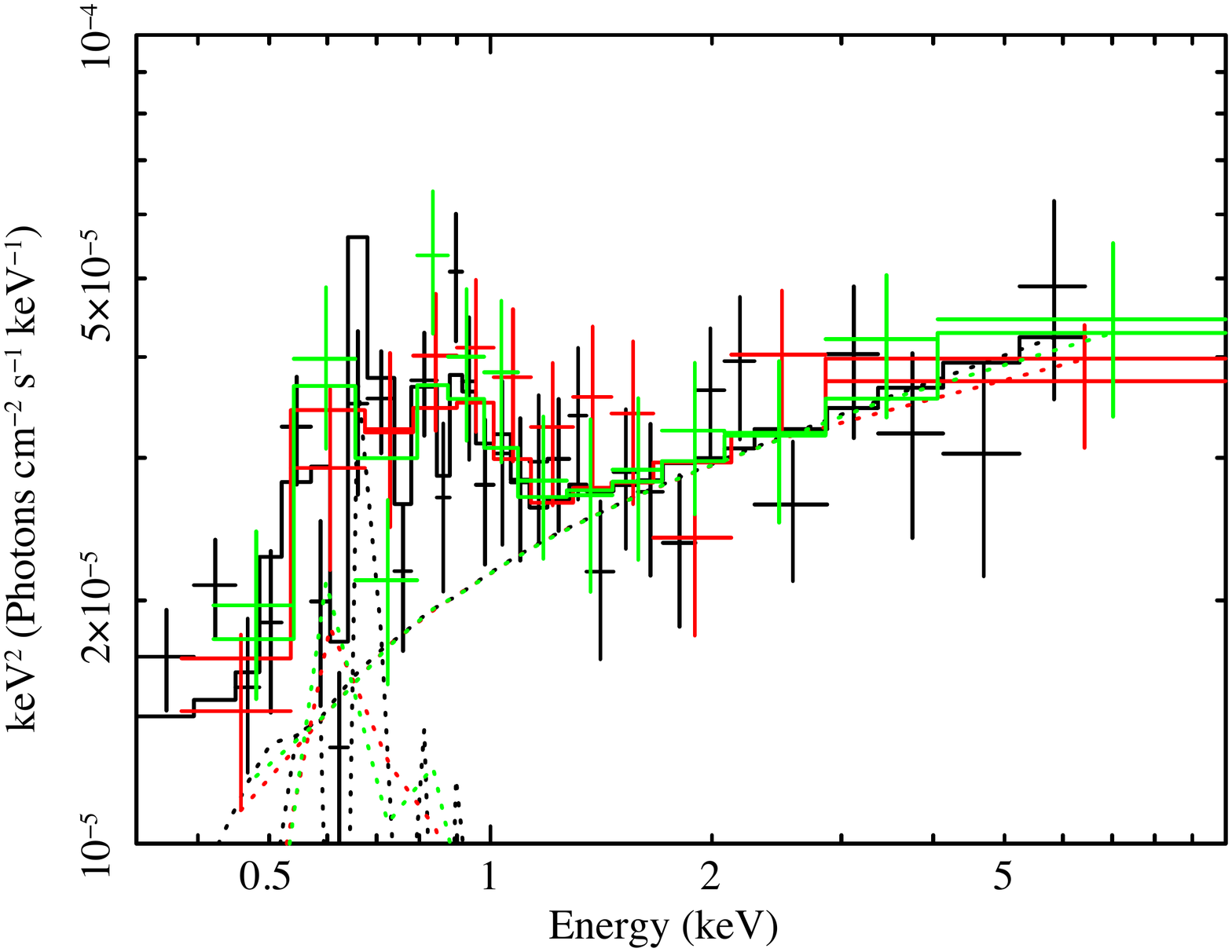}
\hspace{-1.2cm}
\includegraphics[width=9.3cm]{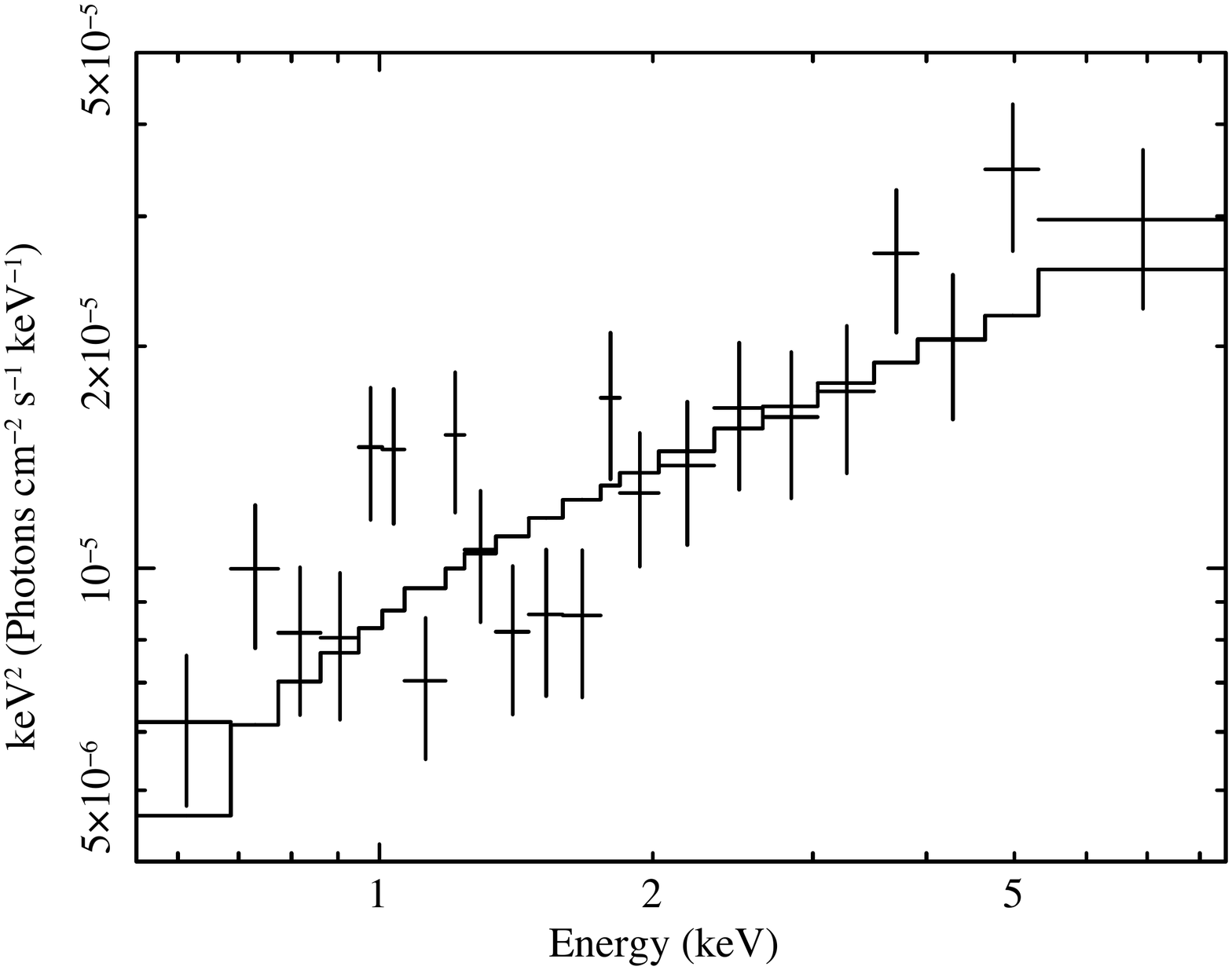}
\end{center}
\vspace{-0.2cm}
\caption{Example spectra of ULX-7 from {\it XMM-Newton} and {\it Chandra} at high and low fluxes, unfolded from the detector response and plotted between 0.3 and 10\,keV, along with the best-fitting absorbed power-law model and the contribution from diffuse emission in the case of {\it XMM-Newton}. Fluxes can be found in Table~\ref{tab:obs}, the diffuse emission parameters in Table~\ref{tab:diff}, and best-fitting source parameters in Table~\ref{tab:source}. For the {\it XMM-Newton} spectra, pn data is plotted in black, MOS1 data in red and MOS2 in green. {\it Top left}, high-flux {\it XMM-Newton} observation X4. {\it Top right}, high-flux {\it Chandra} observation C7. {\it Bottom left}, low-flux {\it XMM-Newton} observation X1. Diffuse emission can be seen to be dominant at the soft end of this spectrum ($<1$\,keV). {\it Bottom right}, low-flux {\it Chandra} observation C10.}
\label{fig:spec}
\end{figure*}

While six {\it XMM-Newton} observations of ULX-7 exist, there is only sufficient data quality for spectral analysis from the first five. We fit the spectra of each of these first five observations with an absorbed power-law model and two additional {\tt mekal} components to account for contamination from the diffuse emission ({\tt mekal+mekal+tbabs*powerlaw}). We set the lower bound of $N_{\rm H}$ for the {\tt tbabs} component to the Galactic foreground value\footnote{Foreground $N_{\rm H}$ was obtained from the HEASARC $N_{\rm H}$ calculator at http://heasarc.gsfc.nasa.gov/cgi-bin/Tools/w3nh/w3nh.pl, using results from \citet{kalberla05}.} of $N_{\rm H} = 1.8 \times 10^{20}$\,cm$^{-2}$ and fixed the {\tt mekal} parameters and normalisations to the average values determined from the {\it Chandra} results, which we take as a good first-order approximation to the contribution of diffuse emission to the spectrum. Most of the {\it XMM-Newton} source spectra are well-fitted by this model and exhibit fairly hard ($\Gamma\sim1.5-1.6$) power-law emission. The exception is observation X3 for which we reject a simple absorbed power-law at $>4\sigma$ significance. The fit for X3 undergoes moderate improvement ($\Delta\chi^2\sim17$ for 2 fewer degrees of freedom) by the addition of a multicolour disc component ({\tt mekal+mekal+tbabs*(diskBB+powerlaw)}), although it is still rejected at $\sim3.5\sigma$ significance. It is unclear from the residuals what a better model might be, so we are unable to find an acceptable fit for the data from this observation. It is possible that ULX-7 exhibits similar soft atomic features to those seen in other ULXs (e.g. \citealt{middleton15b}), however the presence of diffuse emission in the host galaxy complicates more detailed study of the soft end of the spectrum.

We also fit the {\it Chandra} source observations of sufficient data quality with an absorbed power-law model to ensure that they are consistent with the {\it XMM-Newton} results (we do not include the {\tt mekal} components as we assume that the contribution from surrounding diffuse emission is negligible in the {\it Chandra} source data). As in the case of {\it XMM-Newton}, $N_{\rm H}$ is given a lower limit of the Galactic foreground value and allowed to vary, except for observation C10 for which we set $N_{\rm H}$ to $1\times10^{21}$\,cm$^{-2}$ (the average value found from fits to other observations) since there is insufficient data to constrain it further. We find that the spectra are consistent with the same hard ($\Gamma\sim1.5$) power-law shape as the {\it XMM-Newton} observations.

Best fit parameter values for {\it XMM-Newton} and {\it Chandra} are given in Table~\ref{tab:source}, and examples of high- and low-flux spectra and their power-law fits are shown in Fig.~\ref{fig:spec}.

\begin{table}
\caption{The parameter values and goodness of fit for the {\it XMM-Newton} and {\it Chandra} source spectra when fitted with an absorbed power-law model (and a power-law with a multicolour accretion disc in the case of X3).} \label{tab:source}
\begin{center}
\begin{tabular}{ccccc}
  \hline
  ID$^a$ & $N_{\rm H}$ & $\Gamma$ & $T_{\rm in}$ & $\chi^2$/dof \\
   & ($\times10^{21}$ cm$^{-2}$) &  & (keV) & \\
  \hline
  \multicolumn{5}{c}{\textit{\textbf{XMM-Newton}}} \\
  \rule{0pt}{2.5ex}
  X1 & $0.6^{+0.6}_{-0.5}$ & $1.7\pm0.2$ & ... & $47.9/53$ \\
  X2 & $1.1\pm0.2$ & $1.59\pm0.06$ & ... & $231.2/177$ \\
  X3 & $1.1\pm0.2$ & $1.57^{+0.07}_{-0.06}$ & ... & $260.5/174$ \\
   & $1.2\pm0.4$ & $1.2\pm0.2$ & $0.4\pm0.1$ & $243.7/172$ \\
  X4 & $0.8\pm0.2$ & $1.45^{+0.06}_{-0.05}$ & ... & $172.3/182$ \\
  X5 & $0.6^{+0.6}_{-0.5}$ & $1.5^{+0.2}_{-0.1}$ & ... & $30.5/39$ \\
  \multicolumn{5}{c}{\textit{\textbf{Chandra}}} \\
  \rule{0pt}{2.5ex}
  C1 & $1.5^{+0.9}_{-0.8}$ & $1.3\pm0.2$ & ... & $26.2/28$ \\ 
  C3 & $1.4\pm0.4$ & $1.5\pm0.1$ & ...& $67.2/94$ \\
  C5 & $0.4^{+4.0}_{-0.4}$ & $1.3^{+0.5}_{-0.3}$ & ... & $13.7/12$ \\
  C6 & $1.2\pm0.2$ & $1.49\pm0.05$ & ... & $244.1/215$ \\
  C7 & $1.4\pm0.2$ & $1.48^{+0.05}_{-0.04}$ & ... & $273.1/245$ \\
  C8 & $1.6^{+0.3}_{-0.2}$ & $1.54^{+0.06}_{-0.05}$ & ... & $203.6/203$ \\
  C9 & $1.2^{+0.6}_{-0.5}$ & $1.4\pm0.1$ & ... & $65.9/69$ \\
  C10 & $1.0^b$ & $1.5\pm0.2$ & ... & $32.4/20$ \\
  C11 & $1.0\pm0.5$ & $1.5\pm0.1$ & ... & $89.3/94$ \\
  \hline
\end{tabular}
\end{center}
$^a$The short observation ID as defined in Table~\ref{tab:obs}.\\
$^bN_{\rm H}$ frozen at $1\times10^{21}$~cm$^{-2}$.
\end{table}

\begin{figure}
\begin{center}
\includegraphics[width=9cm]{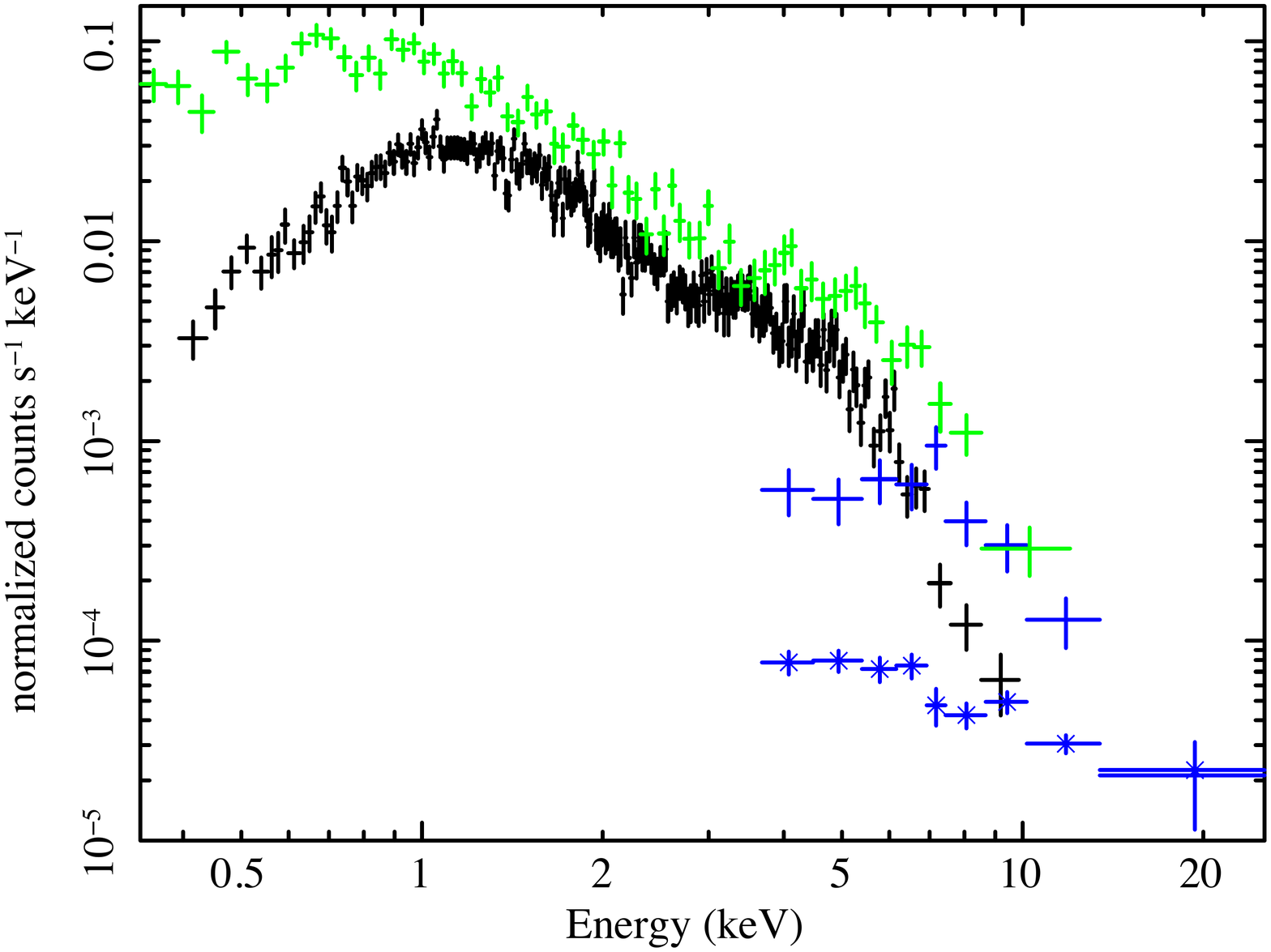}
\end{center}
\caption{The data spectra of the source from {\it XMM-Newton} observation X2 (green -- only EPIC-pn data is shown for clarity), {\it Chandra} observation C6 (black) and combined FPMA and FPMB data from {\it NuSTAR} (blue), along with the {\it NuSTAR} background spectrum (blue crosses). We detect a good signal from {\it NuSTAR} up to $\sim20$\,keV.}
\label{fig:nuback}
\end{figure}

\begin{figure*}
\begin{center}
\includegraphics[width=9.3cm]{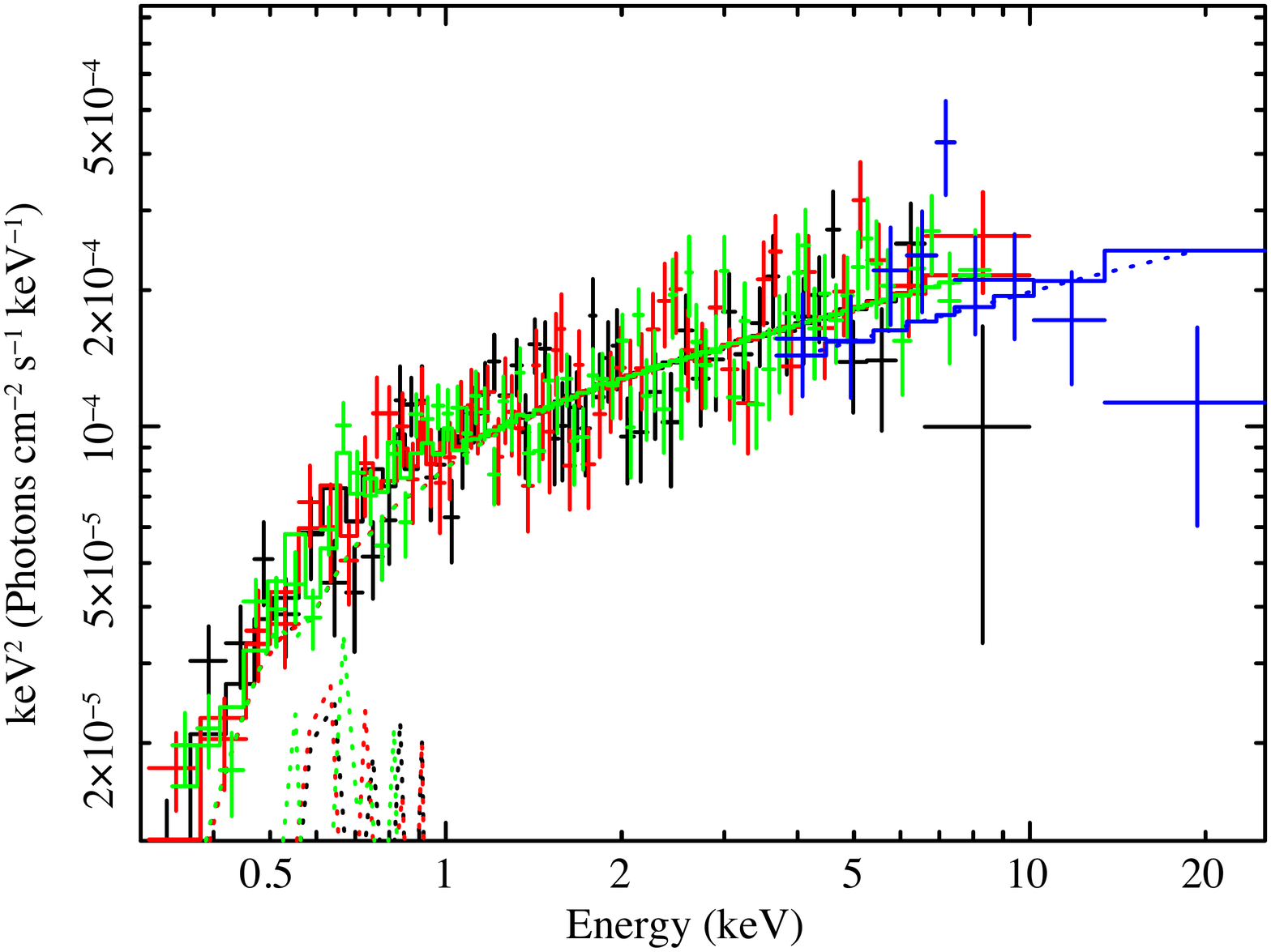}
\hspace{-1.05cm}
\includegraphics[width=9.3cm]{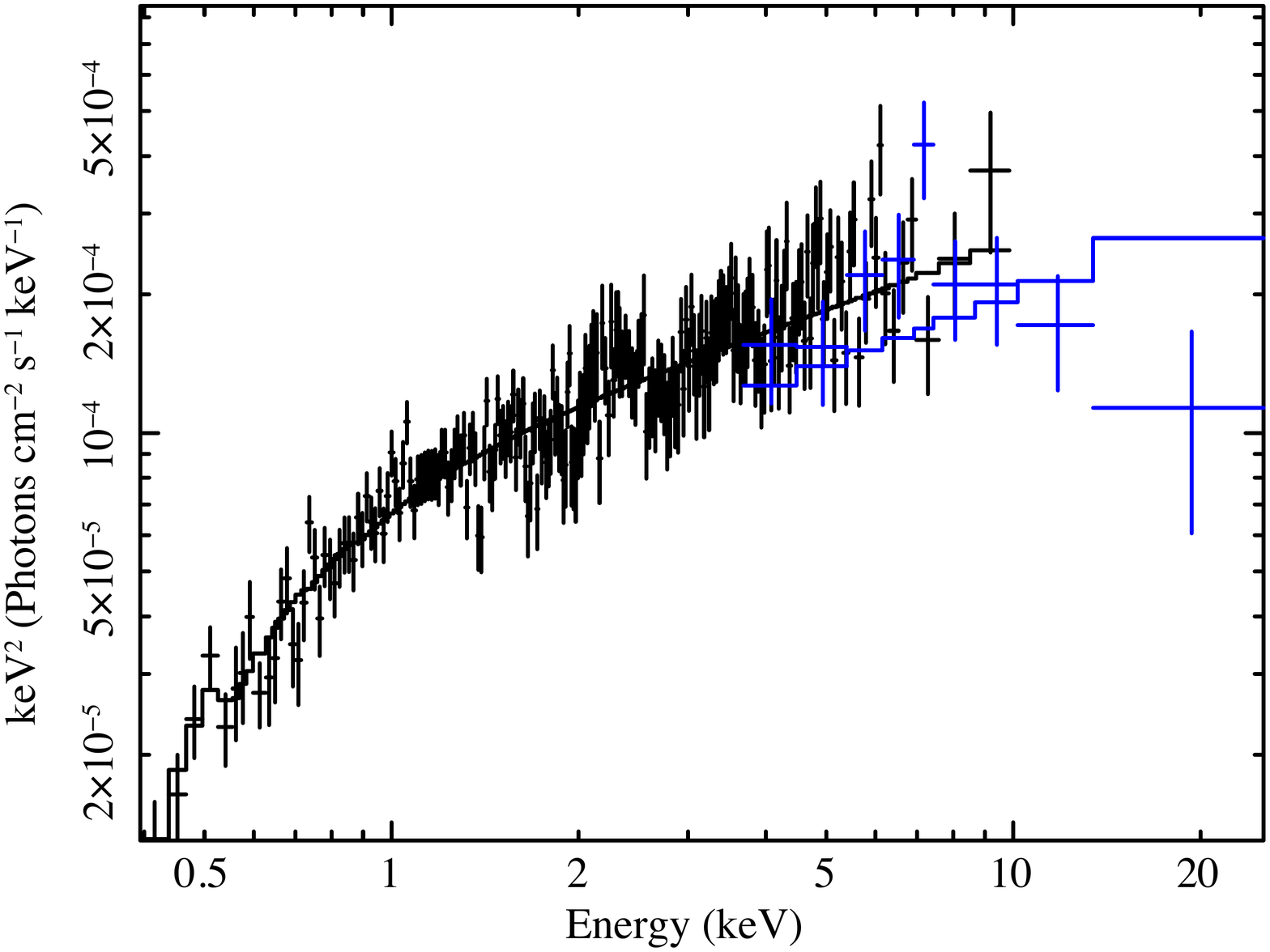}\\
\includegraphics[width=9.3cm]{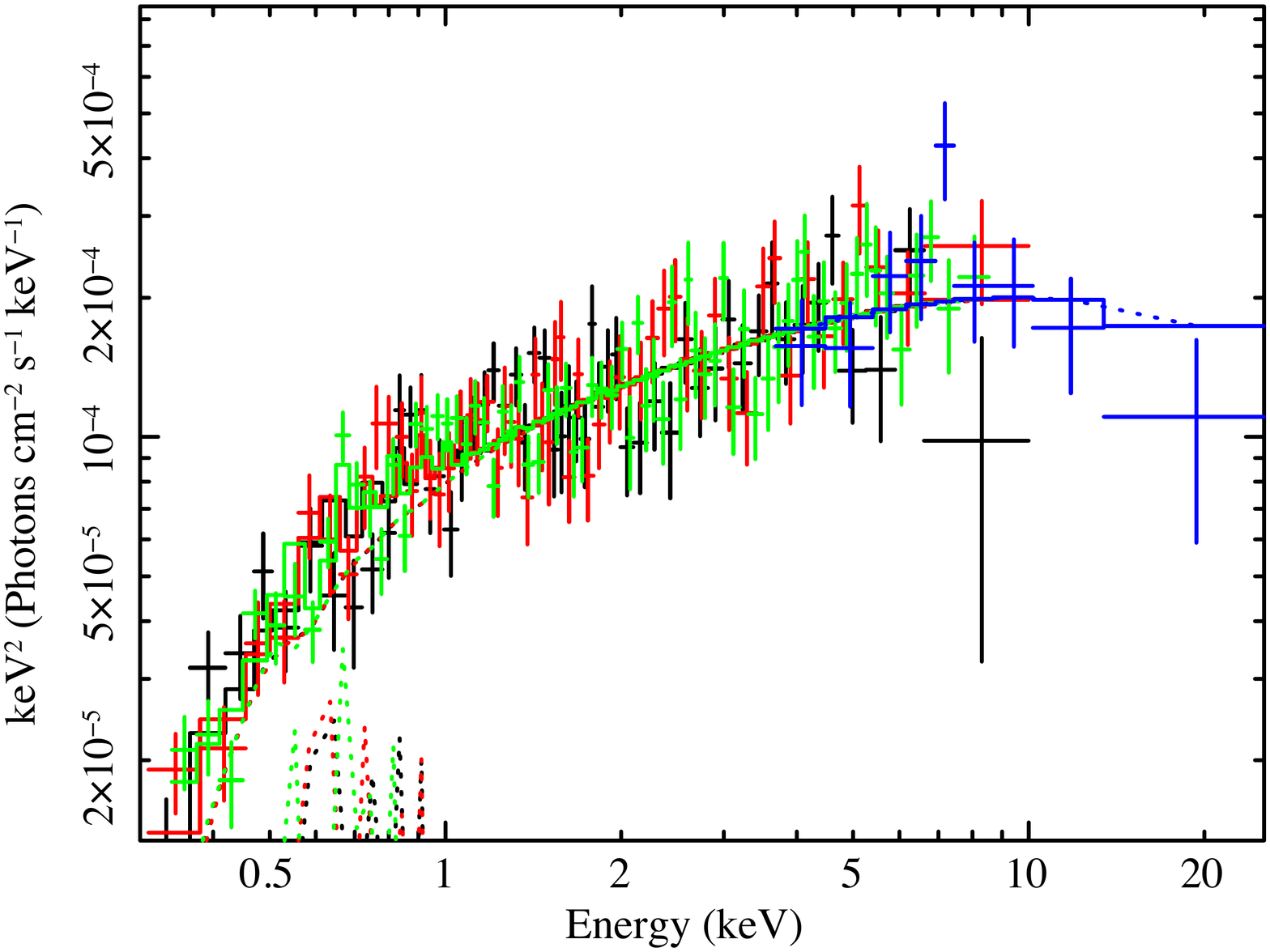}
\hspace{-1.05cm}
\includegraphics[width=9.3cm]{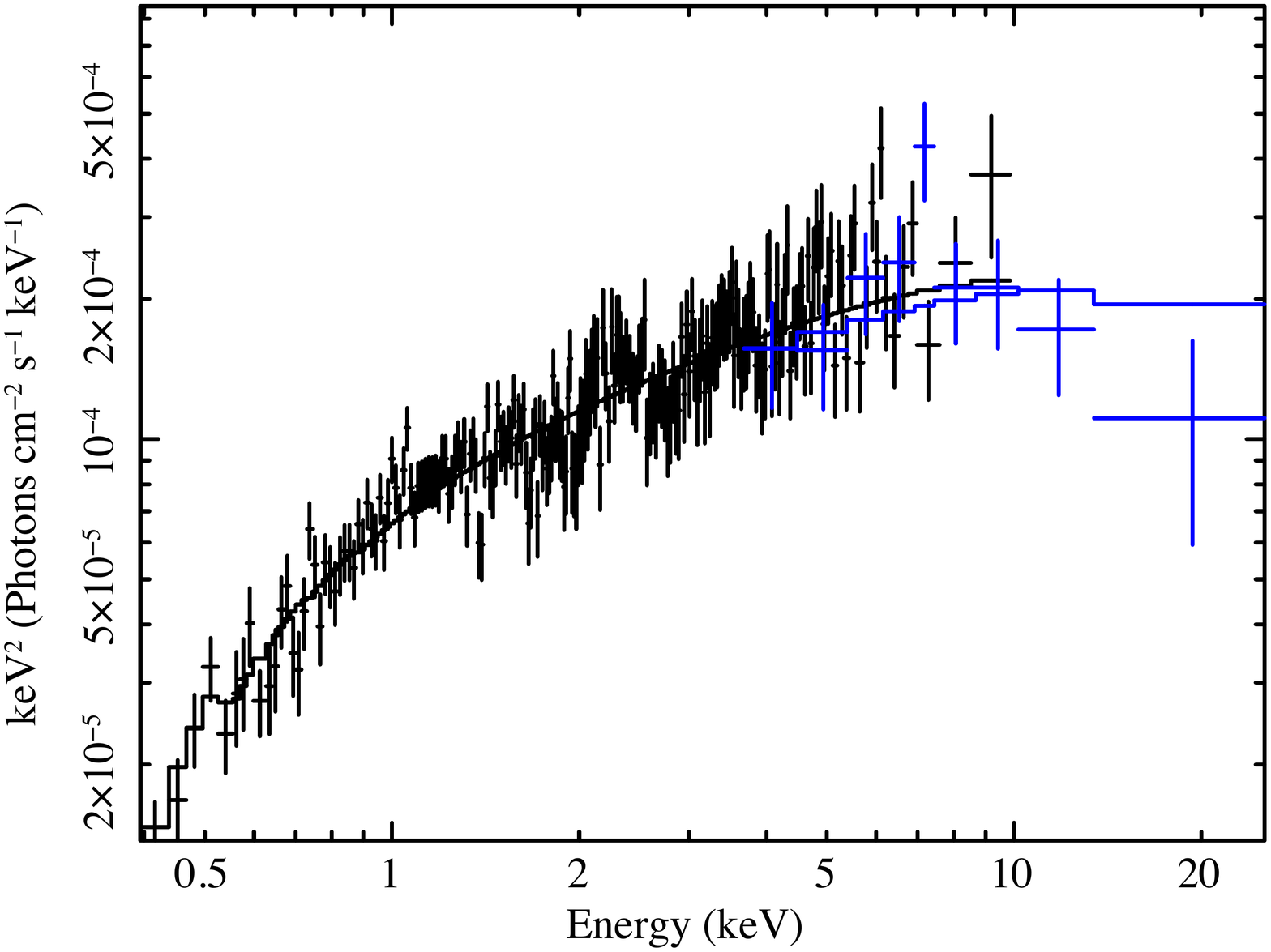}
\end{center}
\vspace{-0.2cm}
\caption{Spectra of ULX-7 from {\it XMM-Newton} observation X2 and {\it Chandra} observation C6, unfolded from the detector response and plotted between 0.3 and 10\,keV, with the combined FPMA and FPMB data from {\it NuSTAR} plotted between 3 and $\sim$20\,keV (blue). {\it XMM-Newton} spectra colours are as Fig.~\ref{fig:spec}. {\it Top left}, X2 and {\it NuSTAR} data, fitted with a {\tt mekal+mekal+tbabs*powerlaw} model. {\it Top right}, C6 and {\it NuSTAR} data, fitted with a {\tt tbabs*powerlaw} model. {\it Bottom left}, X2 and {\it NuSTAR} data, fitted with a {\tt mekal+mekal+tbabs*cutoffpl} model. {\it Bottom right}, C6 and {\it NuSTAR} data, fitted with a {\tt tbabs*cutoffpl} model.}
\label{fig:nuspec}
\end{figure*}

ULX-7 is strongly detected in the 8--24\,keV band in the {\it NuSTAR} observation (Fig.~\ref{fig:image}), with a good signal found up to $\sim20$\,keV (Fig.~\ref{fig:nuback}). The data quality is insufficient to perform in-depth analysis, however we are still able to extract a spectrum from a circular region with radius of 25\,arcseconds. The {\it NuSTAR} spectrum is plotted alongside the {\it XMM-Newton} observation X2 and the {\it Chandra} observation C6 in Fig.~\ref{fig:nuspec}. These are the observations closest in flux to the {\it NuSTAR} observation with the 3--10\,keV flux of the best-fitting absorbed power-law model being $(3.80\pm0.09)\times10^{-13}$ and $(3.9\pm0.2)\times10^{-13}$\,erg\,cm$^{-2}$\,s$^{-1}$ for X2 and C6 respectively. The {\it NuSTAR} 3--10\,keV data has a flux of $(3.8\pm0.7)\times10^{-13}$\,erg\,cm$^{-2}$\,s$^{-1}$ when fitted with an absorbed power-law model, having $\Gamma = 1.3\pm0.7$ (with $N_{\rm H}$ fixed at $1\times10^{21}$\,cm$^{-2}$, the average value found from {\it XMM-Newton} and {\it Chandra} observations, since the {\it NuSTAR} data is too high-energy to constrain it).

The full 3--20\,keV {\it NuSTAR} data demonstrates a softer spectrum when fitted with a power-law model, with $\Gamma = 2.1\pm0.3$. This suggests that the spectrum turns over at the higher energies we observe with {\it NuSTAR}. Therefore we fit the {\it NuSTAR} data simultaneously with observations X2 and C6, with an absorbed power-law model (and a background modelled with {\tt mekal} components for the {\it XMM-Newton} observation as before). We also fit a model replacing the {\tt powerlaw} component with a {\tt cutoffpl} component to characterise any potential turnover. In both cases we also included a multiplicative constant to the absorbed power-law component of the models, which we allowed to vary freely to account for any difference in normalisation between {\it NuSTAR} and the other telescope. The fit parameters with {\it NuSTAR} data included are given in Table~\ref{tab:nustar}, although it is important to note that neither of these observations are contemporaneous with the {\it NuSTAR} observation and so we cannot be certain that the source is in the same spectral state between them.

While both observations are consistent with a cut-off to the energy spectrum at 18\,keV, we find that a cut-off power-law model offers no improvement over a power-law model for either observation X2 or C6. This is not entirely unexpected, as the appearance of a turnover is mainly driven by a single {\it NuSTAR} data point. Further observations with {\it NuSTAR} simultaneous with observations from {\it XMM-Newton} are required to better constrain the high-energy spectral shape of this source.

\begin{table}
\caption{Parameter values and goodness of fit for the {\it NuSTAR} spectrum fit simultaneously with the closest-flux observations X2 and C6, with both a power-law (top) and a cut-off power-law (bottom) model.} \label{tab:nustar}
\vspace{-0.2cm}
\begin{center}
\begin{tabular}{cccccc}
  \hline
  \hline
  ID & $N_{\rm H}$ & $\Gamma$ & $E_{\rm cut}$ & $\chi^2$/dof \\
   & ($\times10^{21}$ cm$^{-2}$) &  & (keV) & \\
  \hline
  X2 & $1.4\pm0.3$ & $1.64\pm0.06$ & ... & $245.0/184$ \\
   & $1.1\pm0.3$ & $1.5\pm0.1$ & $18^{+51}_{-8}$ & $239.9/183$ \\
  C6 & $1.3\pm0.2$ & $1.51\pm0.05$ & ... & $264.5/223$ \\
   & $1.1\pm0.3$ & $1.3\pm0.1$ & $18^{+43}_{-8}$ & $258.8/222$ \\
  \hline
\end{tabular}
\end{center}
\end{table}

\subsection{Timing Analysis}
\label{sec:time}

All observations of ULX-7 with {\it XMM-Newton} are flagged as variable in the {\it XMM-Newton} Serendipitous Source Catalogue, and all have fractional rms at $\sim$30-40\% according to an initial examination of the light curves using the {\sc lcstats} routine in {\sc FTOOLS}. Previous studies have attempted to find a period in this variability, with \citet{liu2002} suggesting a period of 7620\,s using {\sc efsearch} and \citet{dewangan05} similarly declaring a period of 5925\,s with $\sim2\sigma$ significance. However, a subsequent study by \citet{terashima06} found no evidence of periodic variation, instead suggesting that the source variability is due to stochastic noise. 

\begin{figure}
\begin{center}
\includegraphics[width=9cm]{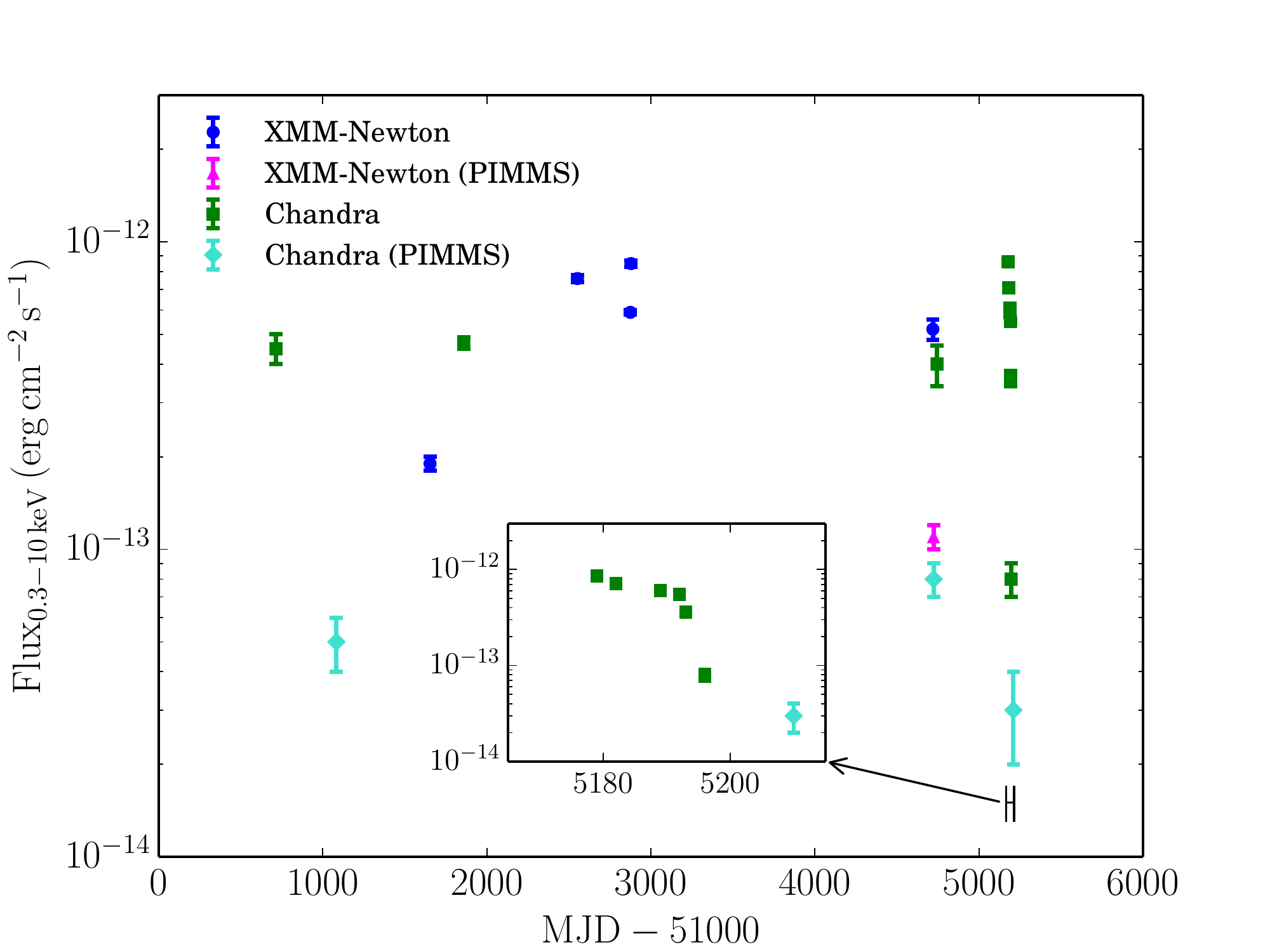}
\includegraphics[width=9cm]{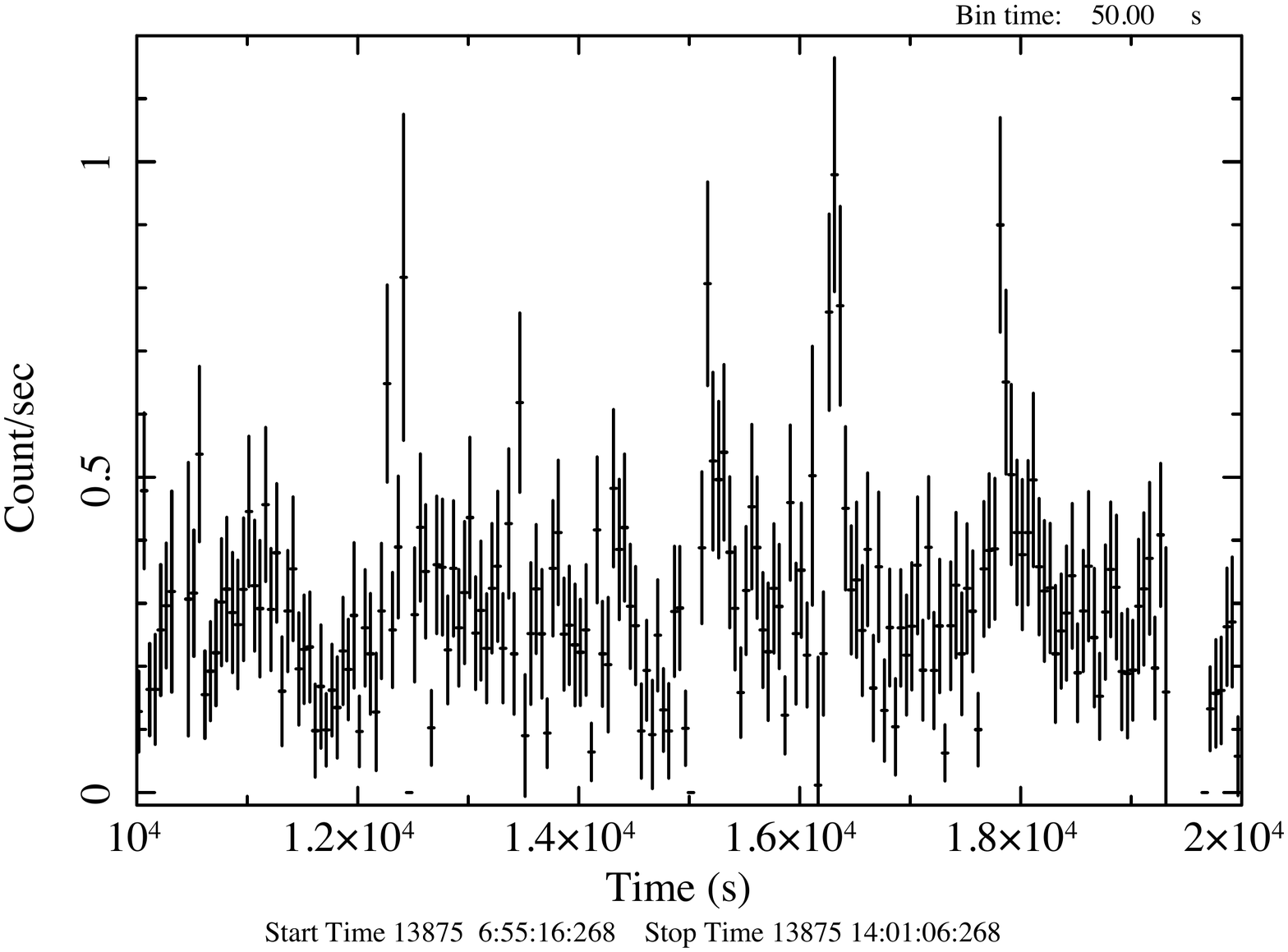}
\end{center}
\vspace{-0.2cm}
\caption{{\it Top}, the long-term lightcurve for ULX-7 showing the 0.3--10\,keV flux over time as measured using the {\it XMM-Newton} and {\it Chandra} telescopes. Points identified as {\sc PIMMS} are fluxes calculated from the count rate assuming a power-law spectrum with $N_{\rm H} = 1\times10^{21}$\,cm$^{-2}$ and $\Gamma=1.5$. See Table~\ref{tab:obs} for details and values. {\it Bottom}, a 10\,ks segment of the pn + MOS light curve of observation X3, binned into 50\,s intervals.}
\label{fig:lc}
\end{figure}

The source also undergoes significant long-term variation, with the dynamic range of its flux encompassing well over an order of magnitude, even over the course of a single month when observed using {\it Chandra} in 2012. The long-term lightcurve, along with an example of short-term variability from observation X3, is shown in Fig.~\ref{fig:lc}. However, despite this variation in flux there is no evidence for a flux-hardness relation, given the consistent shape of the spectrum found in Section~\ref{sec:spec}.

We created power spectra for the {\it XMM-Newton} and {\it Chandra} observations of ULX-7 by taking the periodogram of fixed-length segments in each observation taken from good time intervals and averaging over all segments for each telescope. We used 3200\,s segments for the {\it XMM-Newton} observations and 12800\,s segments for the {\it Chandra} observations. The greater length of the {\it Chandra} observations allows us to probe down to $\sim10^{-4}$\,Hz, although at higher frequencies the data is dominated by noise, whereas the {\it XMM-Newton} data, while not having the low-frequency range, has far less contribution from white noise up to $\sim8\times10^{-3}$\,Hz. The two datasets are therefore very complementary and allow us access to two decades of frequency space.

The power spectra are normalised so that the power is given in units of the squared fractional rms per frequency interval. We combined all observations for each telescope, given that the overall shape of the power spectrum remained consistent from observation to observation, except for {\it Chandra} observations C4 and C5, which did not have good time intervals long enough for our chosen segment length, and C12, which contributed a lot of noise to the power spectrum due to a very low count rate.

\begin{figure}
\begin{center}
\includegraphics[width=9cm]{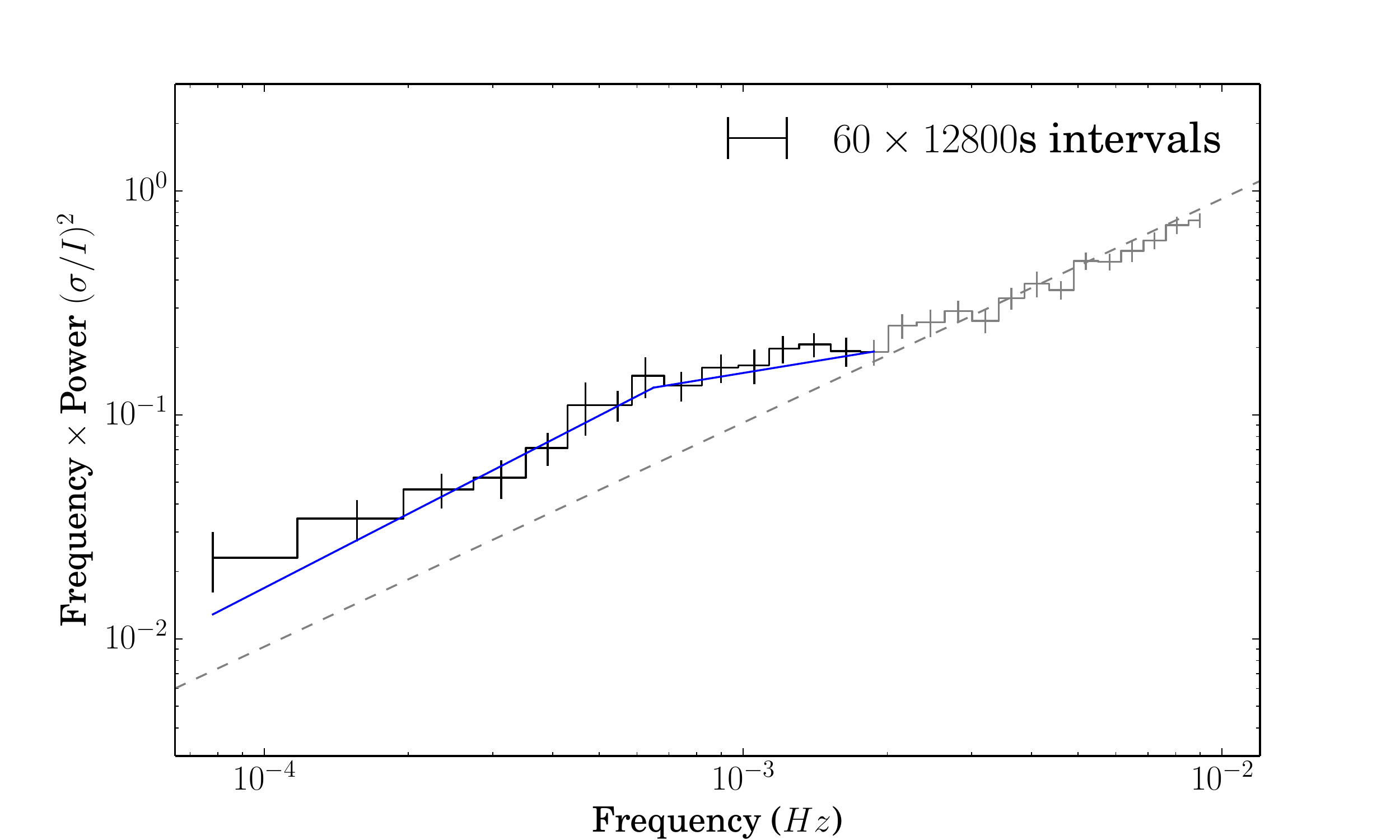}
\includegraphics[width=9cm]{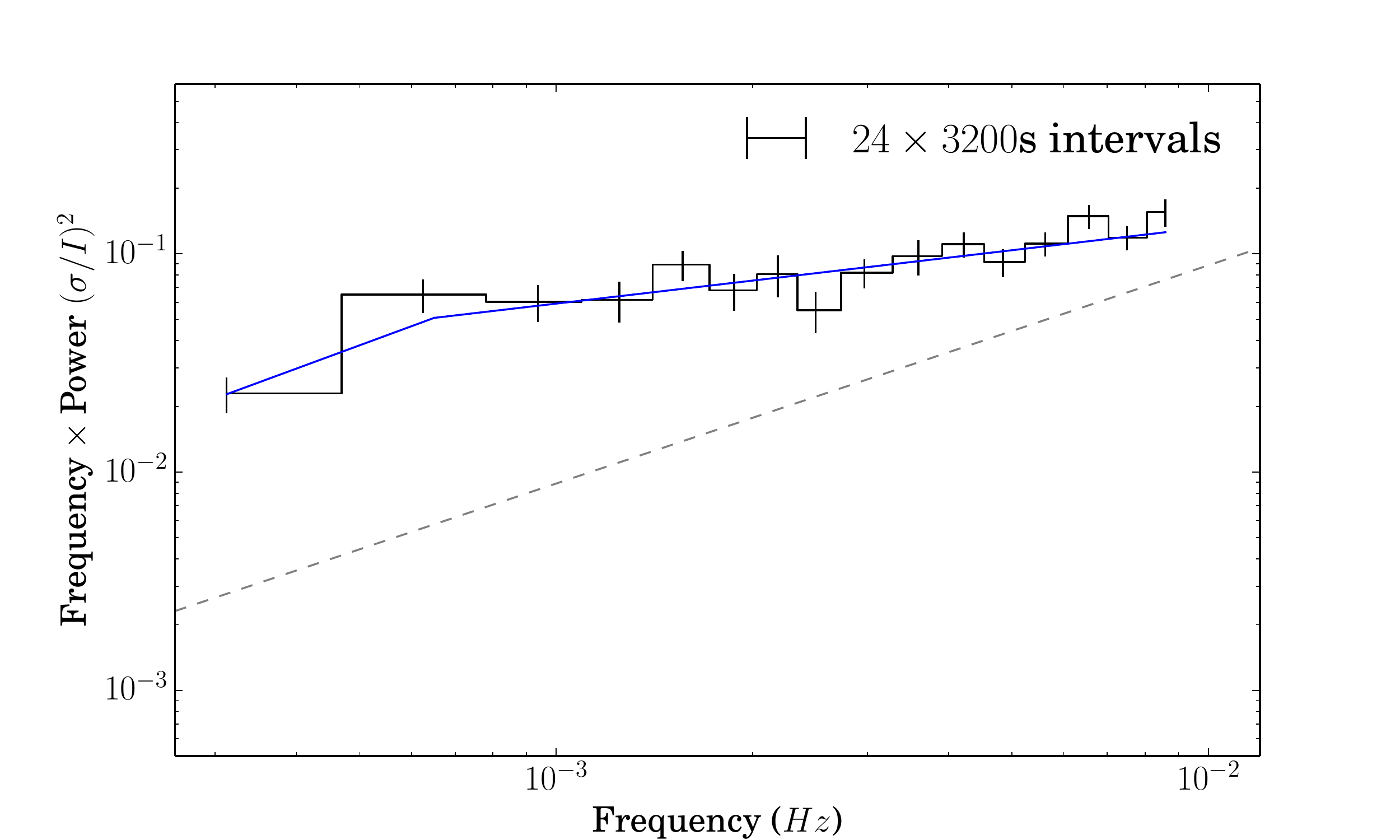}
\end{center}
\caption{The power spectra for {\it Chandra} (top) and {\it XMM-Newton} (bottom) observations, along with the best-fitting broken power-law model in blue (for which $\nu_b=6.5\times10^{-4}$~Hz, $\alpha_1=-0.1$ and $\alpha_2=0.65$), normalised so that power is in units of the squared fractional rms per frequency interval. The white noise level is marked by the dashed grey line, and the bins discounted from analysis due to being dominated by white noise are also coloured grey. The data is geometrically rebinned with a co-efficient of 1.1, and error bars represent the standard error on the mean for each frequency bin.}
\label{fig:powspec}
\end{figure}

We first rule out a simple power-law shape to the power spectrum by performing a simultaneous fit to the {\it XMM-Newton} and {\it Chandra} data using the {\tt whittle} statistic in XSPEC and a {\tt powerlaw} model, disregarding frequency bins consistent with the white noise level of the power spectrum. The best-fitting power-law has $\alpha=0.4$ (for $P(\nu) \propto \nu^{-\alpha}$) which has $\chi^2=54.8/28$ and we reject at $>3\sigma$ significance, so we can be confident that the power spectrum shape requires a more complex model to fit it. We next fit the power spectrum with a broken power-law model. This is an excellent fit to the data, with goodness-of-fit $\chi^2=23.0/25$, however the fit parameters are not highly constrained. We find that the power spectrum exhibits a break at $\nu_b=6.5_{-1.1}^{+0.5}\times10^{-4}$~Hz, with a low-frequency slope of $\alpha_1=-0.1^{+0.5}_{-0.2}$ and a high-frequency slope of $\alpha_2=0.65^{+0.05}_{-0.14}$ (errors were found using a Monte Carlo Markov Chain method with a chain length of 100,000). Since the break is at the overlap of the two power spectra and its frequency can only be constrained with the {\it Chandra} data, it is likely that future long observations with {\it XMM-Newton} will help to better characterise the break. The power spectra for the two telescopes are shown in Fig.~\ref{fig:powspec}.

\begin{figure}
\begin{center}
\includegraphics[width=9cm]{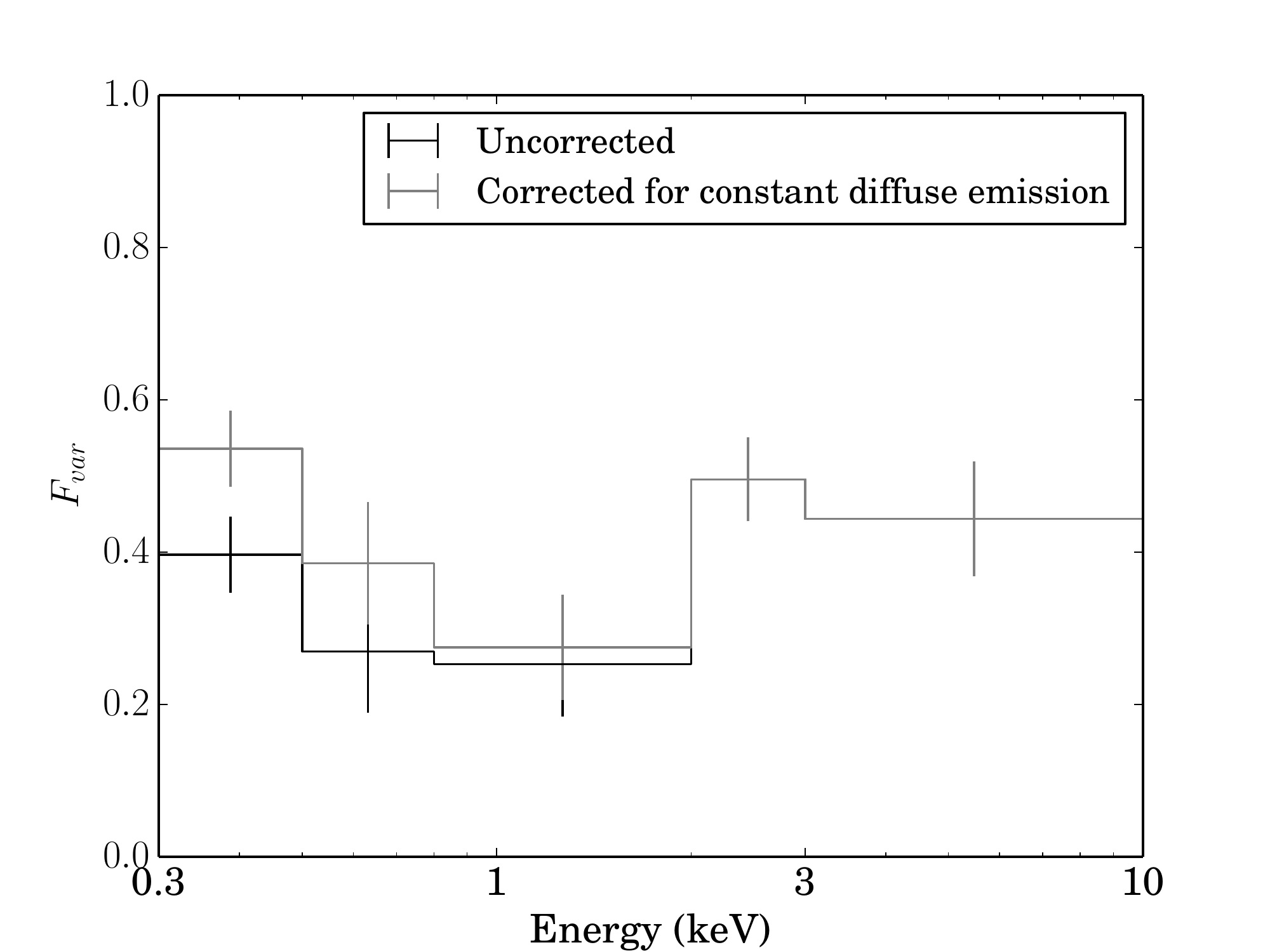}
\end{center}
\caption{The fractional rms spectrum for ULX-7, both uncorrected (black) and corrected to discount the contribution from non-variable diffuse emission (grey). The energy bands are: $0.3-0.5$\,keV, $0.5-0.8$\,keV, $0.8-2.0$\,keV, $2.0-3.0$\,keV and $3.0-10.0$\,keV. The error bars represent the standard error on the mean across all 19 3200\,s segments.}
\label{fig:fracvar}
\end{figure}

In order to see how the fractional variability of ULX-7 changes as a function of energy, we created a fractional rms spectrum using five energy bands by integrating over the power spectrum for each energy band, averaging over all {\it XMM-Newton} segments. Since the source flux is contaminated by diffuse emission that we do not expect to be variable, we also correct for the flux contribution from the diffuse emission, giving the intrisic fractional variability of the source. The source exhibits a high amount of variability across all energy bands, especially at low and high energies, although we find the spectrum to be consistent with constant fractional rms at all energies. The fractional rms spectrum is shown in Fig.~\ref{fig:fracvar}.

\begin{figure}
\begin{center}
\includegraphics[width=9cm]{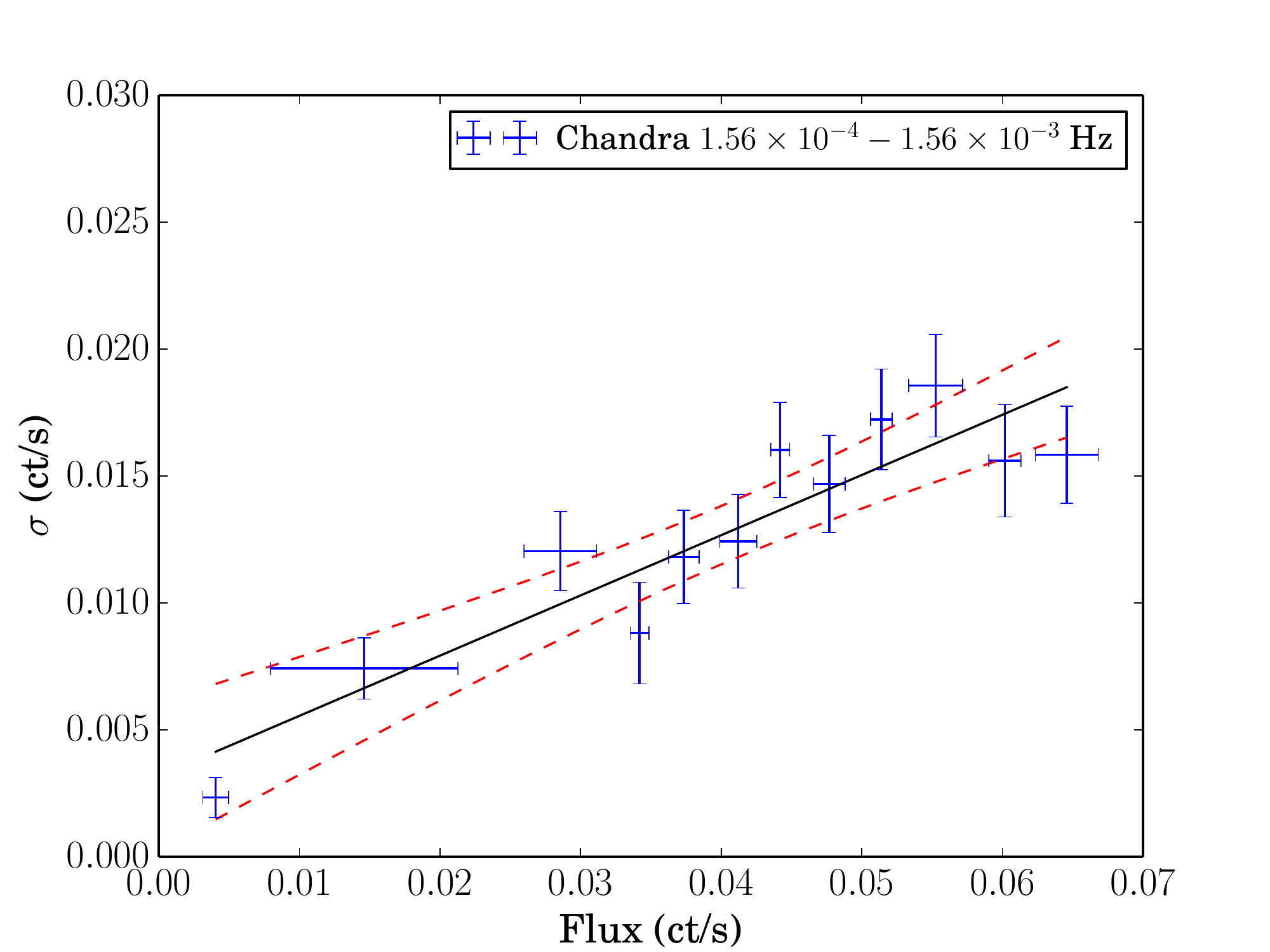}
\includegraphics[width=9cm]{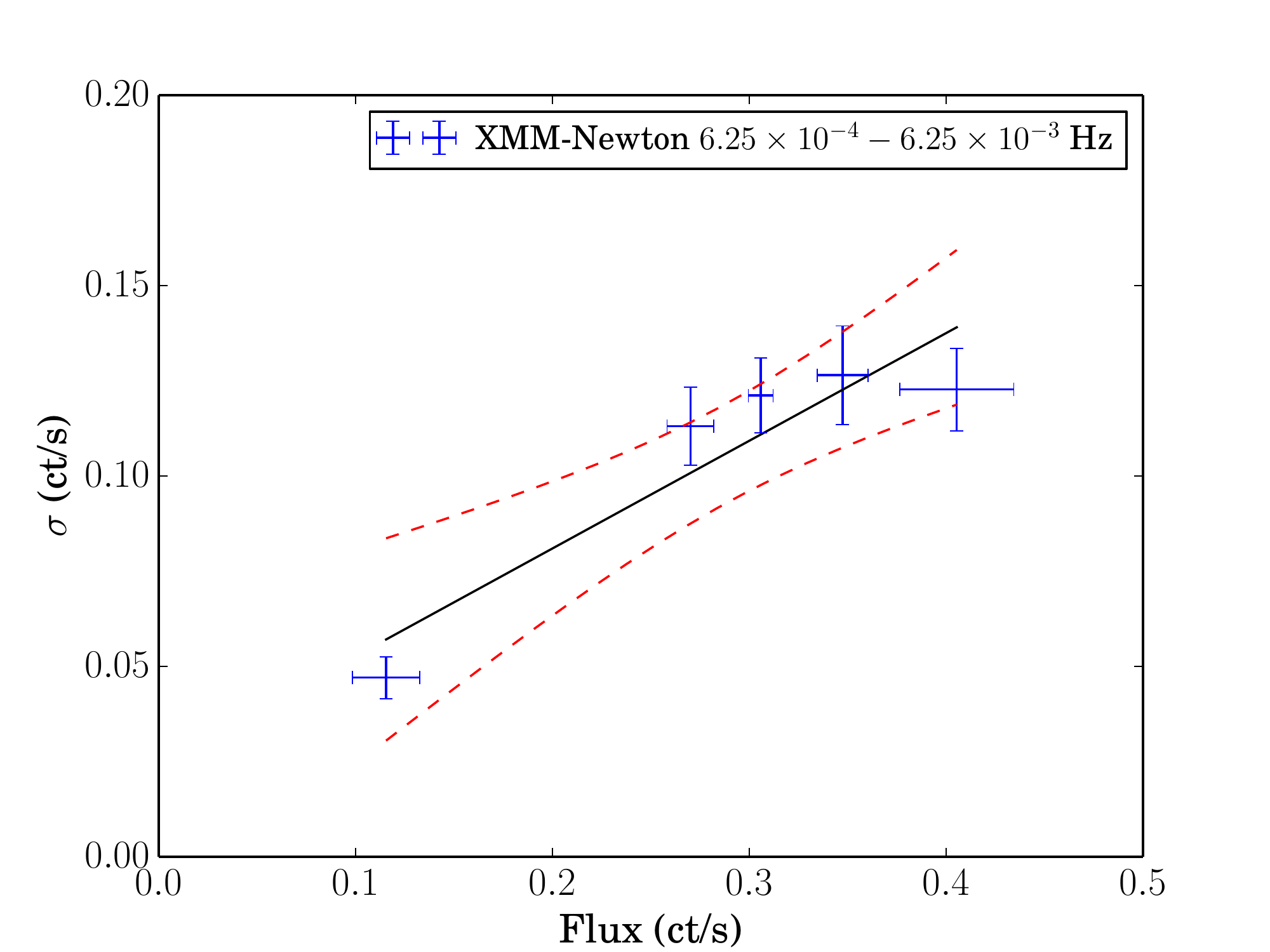}
\end{center}
\caption{The rms-flux relation for {\it Chandra} (top) and {\it XMM-Newton} (bottom) observations, along with a line of best fit (black) and 90\% confidence intervals (red dashed). The error bars are the standard deviation of the rms and flux.}
\label{fig:rmsflux}
\end{figure}

We also checked for an rms-flux relation by ordering intervals by flux and grouping them into bins of at least 10 before creating an unnormalised power spectrum for each bin and integrating over a decade in frequency to find the rms. The data was sufficient to confirm that ULX-7 exhibits a positive linear rms-flux relation as expected for an accreting source \citep{heil12}, with a significance of $>10\sigma$ for a positive slope. The rms-flux relations are shown in Fig.~\ref{fig:rmsflux}. ULX-7 is the third ULX to date for which a positive linear rms-flux relation has been confirmed \citep{hernandezgarcia15}.

Finally, we also examine the data from the {\it XMM-Newton} EPIC-pn camera, which has a time resolution of 73.4\,ms in full-frame mode, for evidence of coherent pulsations such as those produced by pulsars. To do this we use the $H$-test \citep{dejager89}. In brief, the $H$-test is a test for a periodic signal that is especially useful in the case where there is no a priori information about the shape of the light curve available. The $H$ statistic is based on the $Z_m^2$ statistic \citep{buccheri83}, and defines the optimal number of harmonics, $M$, such that:

\begin{center}
\begin{equation}
H \equiv \max_{1 \leq m \leq 20} \big( Z^2_m + 4m + 4 \big) = Z^2_M + 4M + 4 \geq 0
\label{eq:htest}
\end{equation}
\end{center}

We apply the $H$-test to the five longest {\it XMM-Newton} observations, examining a range of frequencies from 6.85\,Hz (approximately the Nyquist frequency for EPIC-pn data) to 0.1\,Hz. We found no evidence of a pulsation period to high significance (that is, the commonly quoted condition of $H>23$), although we found three marginally significant periods with $H>17$, equivalent to a probability of $p \lesssim 10^{-3}$ that these $H$ values were produced by chance, which are listed in Table~\ref{tab:periods}. It should be noted that a period of 0.3003\,s is approximately 4 times the time resolution of the EPIC-pn camera.

\begin{table}
\caption{Periods with $p<10^{-3}$ found in {\it XMM-Newton} EPIC-pn data when using the $H$-test to search for pulsations.} \label{tab:periods}
\begin{center}
\begin{tabular}{ccc}
  \hline
  \hline
  ID$^a$ & Period & $p^b$ \\
   & (s) & ($\times 10^{-4}$) \\
  \hline
  X2 & 0.1833 & 9.62 \\
  X2 & 0.3003 & 4.71 \\
  X3 & 0.6211 & 2.60 \\
  \hline
\end{tabular}
\end{center}
$^a$The short observation ID as defined in Table~\ref{tab:obs}.\\
$^b$The probability of the null hypothesis that there is no periodic signal for this period.
\end{table}

\subsection{Optical Counterparts}
\label{sec:hubble}

\begin{figure}
\begin{center}
\includegraphics[width=7cm]{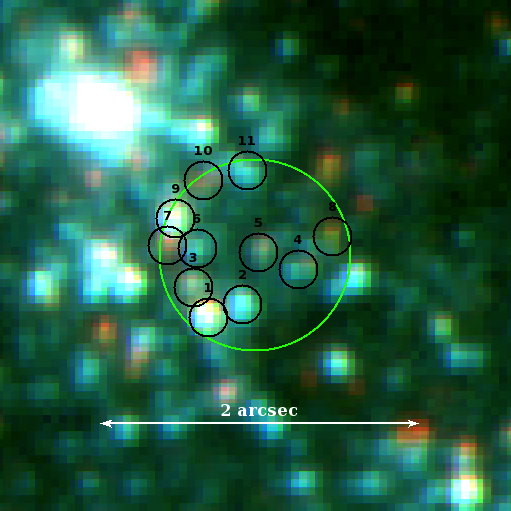}
\end{center}
\caption{{\it HST} true-colour image around the position of ULX-7, with the red, green and blue channels corresponding to the F435W, F555W and F814W bands respectively. A 0.6\,arcsecond radius circle of 90\% confidence is shown around the source position 13:30:01.0~+47:13:44. The numbered objects correspond to those listed in Table~\ref{tab:stars}.}
\label{fig:hst}
\end{figure}

We mark the {\it Chandra} position of ULX-7 on a true color {\it HST} image with a 0.6\,arcsecond radius 90\% confidence circle in Fig.~\ref{fig:hst}. Using {\sc DAOPHOT II}, we were able to obtain photometric data for 11 objects within the circle, although visual inspection reveals that there are other possible counterparts that are too faint or unresolved to characterise. A list of objects and their magnitudes as determined by {\sc DAOPHOT II} is given in Table~\ref{tab:stars}. Given the faint and crowded nature of the field, we do not expect the values we obtain to be more than approximations.

\begin{figure}
\begin{center}
\includegraphics[width=8cm]{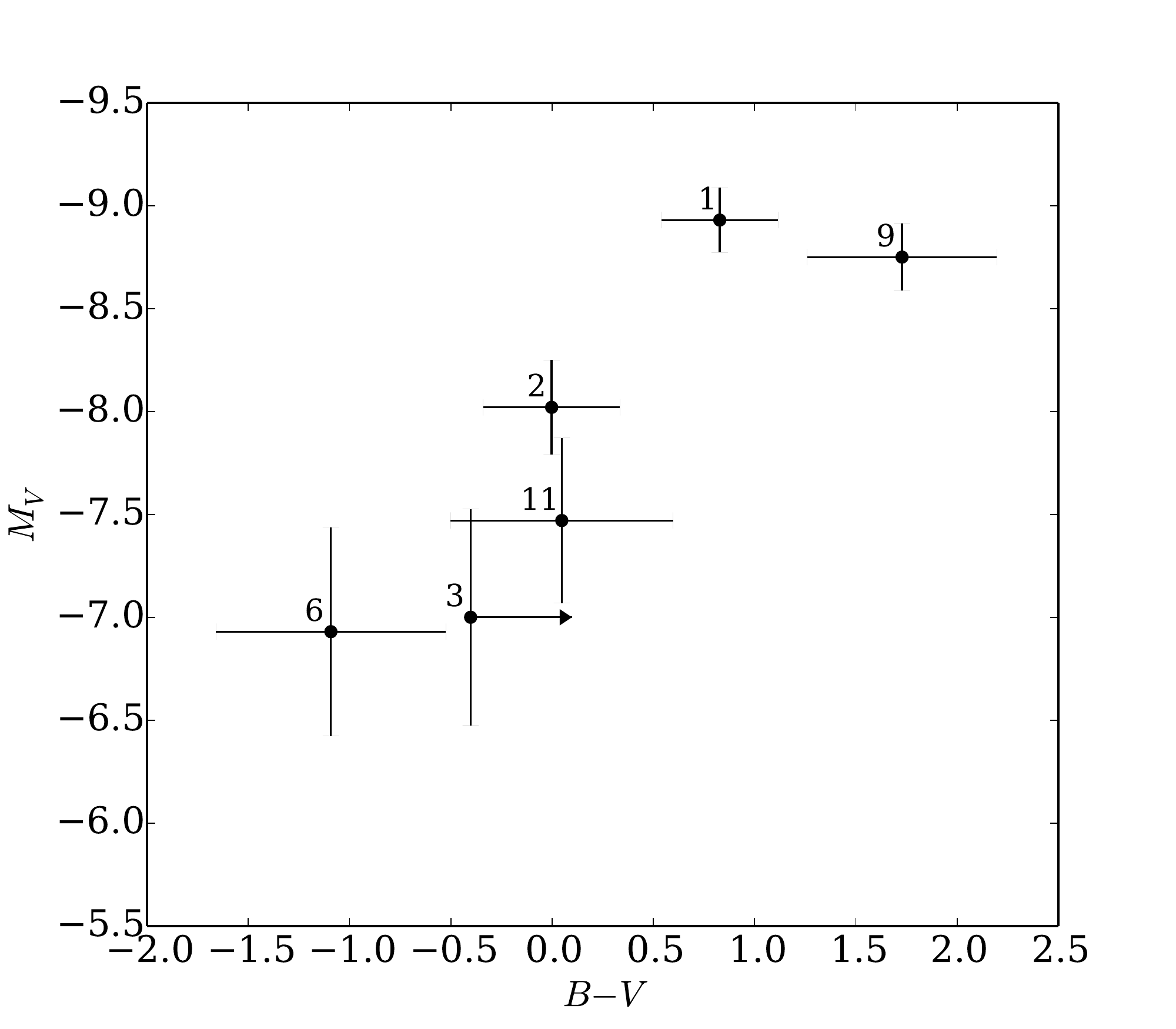}
\end{center}
\caption{$B-V$ colour against the $V$-band magnitude for 6 of the 11 potential optical counterparts within 0.6\,arcseconds of ULX-7, for which we were able to calculate a magnitude for the $B$- and/or $V$-band. Numbers correspond to the ID column of Table~\ref{tab:stars}. Errors are the estimated standard error from {\sc DAOPHOT II} results.}
\label{fig:colmag}
\end{figure}

Using the distance modulus for M51, $\mu=29.45$, we can calculate an absolute magnitude for each object. We plot $M_V$ against the $B-V$ colour in Fig.~\ref{fig:colmag}. Most of the objects that we are able to characterise have low absolute magnitudes (that is, high luminosities) and $B-V$ colours consistent with OB supergiants \citep{roberts08} -- suitable companion stars for a HMXB -- for which we would expect values of $M_V$ of between $-7$ and $-4$, and $B-V\sim-0.2$ \citep{wegner06, roberts08}. This appears to be consistent with previous findings indicating that ULX optical counterparts are often consistent with being OB-type stars (e.g. \citealt{gladstone13}). We might also expect these properties from an X-ray irradiated disc (e.g. \citealt{madhusudhan08, tao11}). The two exceptions are objects 1 and 9, which are significantly brighter and redder than expected for a OB-type star, and too luminous to be red supergiants (for which we would expect $M_V\sim-6$; \citealt{heida14}). Because of this, it is likely that objects 1 and 9 are small, unresolved clusters of multiple stars. This may also be the case for object 2, as it is also unusually bright.

\begin{table*}
\begin{minipage}{164mm}
\caption{The positions, magnitudes and colours of optical counterparts found within a 0.6\,arcsecond error circle of 13:30:01.0~+47:13:44, characterised using {\sc DAOPHOT II}.} \label{tab:stars}
\begin{tabular}{ccccccccccc}
  \hline
  \hline
  ID$^a$ & R.A. (J200) & Dec. (J200) & $m_B^b$ & $m_V^b$ & $m_I^b$ & $M_V^c$ & $B-V^c$ & $V-I^c$ & $F_{\rm X}/F_{\rm opt}^d$ \\
  \hline
  1 & 13~30~01.05 & 47~13~43.62 & $22.1\pm0.2$ 	& $21.43\pm0.09$ & $21.04\pm0.07$ & $-8.9\pm0.2$ & $0.8\pm0.3$ 	& $0.5\pm0.2$ 	& $11.7\pm0.4$\\
  2 & 13~30~01.03 & 47~13~43.70 & $22.2\pm0.2$ 	& $22.3\pm0.2$ 	& $22.8\pm0.3$ 	& $-8.0\pm0.2$ 	& $0.0\pm0.3$ 	& $-0.4\pm0.4$ 	& $27.0\pm0.9$\\
  3 & 13~30~01.06 & 47~13~43.80 & $>22.8$ 		& $23.4\pm0.5$ 	& $22.6\pm0.3$ 	& $-7.0\pm0.5$ 	& $>-0.4$ 		& $0.8\pm0.6$ 	& $69\pm2$ \\
  4 & 13~30~00.99 & 47~13~43.91 & $>22.8$ 		& $>23.5$ 		& $23.3\pm0.4$ 	& $>-6.9$ 		& ...			& $>0.3$ 		& $>76$ \\
  5 & 13~30~01.02 & 47~13~44.03 & $>22.8$ 		& $>23.5$ 		& $23.0\pm0.3$ 	& $>-6.9$ 		& ...			& $>0.6$ 		& $>76$ \\
  6 & 13~30~01.06 & 47~13~44.04 & $22.2\pm0.2$ 	& $23.4\pm0.5$ 	& $>23.9$ 		& $-6.9\pm0.5$ 	& $-1.1\pm0.5$ 	& $<-0.4$ 		& $73\pm3$ \\
  7 & 13~30~01.07 & 47~13~44.07 & $>22.8$ 		& $>23.5$ 		& $22.2\pm0.2$ 	& $>-6.9$ 		& ... 			& $>1.3$ 		& $>76$ \\
  8 & 13~30~00.97 & 47~13~44.12 & $>22.8$ 		& $>23.5$ 		& $23.0\pm0.4$ 	& $>-6.9$ 		& ... 			& $>0.5$ 		& $>76$ \\
  9 & 13~30~01.07 & 47~13~44.23 & $23.2\pm0.4$ 	& $21.6\pm0.1$ 	& $21.5\pm0.1$ 	& $-8.8\pm0.2$ 	& $1.7\pm0.5$ 	& $0.3\pm0.3$ 	& $13.8\pm0.5$ \\
  10 & 13~30~01.05 & 47~13~44.47 & $>22.8$		& $>23.5$ 		& $22.7\pm0.3$ 	& $>-6.9$ 		& ...			& $>0.8$ 		& $>76$ \\
  11 & 13~30~01.03 & 47~13~44.52 & $22.8\pm0.3$ 	& $22.9\pm0.4$ 	& $>23.9$ 		& $-7.5\pm0.4$ 	& $0.0\pm0.5$ 	& $<-0.9$ 		& $45\pm2$ \\
  \hline
\end{tabular}
$^a$ID for the purposes of reference within this paper only.\\
$^b$Observed magnitude and estimated standard error from {\sc DAOPHOT II} results, or lower limits where the source was able to be characterised with {\sc DAOPHOT II}.\\
$^c$Absolute magnitude and colours, corrected for foreground extinction using $E(B-V) = 0.0301\pm0.0007$, found using the IRSA online calculator for Galactic dust and reddening (http://irsa.ipac.caltech.edu/applications/DUST). Values used are from \citet{schlafly11}.\\
$^d$X-ray/optical flux ratio based on the highest detected 2--10\,keV flux from ULX-7. 
\end{minipage}
\end{table*}

While the colours for the rest of the objects are consistent with OB-type stars, we would also expect background quasars at intermediate redshifts to appear blue, and it would be reasonable to detect them at similar apparent magnitudes to these objects. Therefore we also calculated an X-ray/optical ratio for the objects, using the highest 2--10\,keV flux recorded from ULX-7 and calculating the optical flux using the formula $F_{\rm opt}=8\times10^{-6}\cdot10^{-m_V/2.5}$\,erg\,cm$^{-2}$\,s$^{-1}$ (e.g. \citealt{shtykovskiy05}). These ratios are given in Table~\ref{tab:stars}. For any potential counterpart within the error circle that is not characterised by DAOPHOT II, including the faint red objects for which we were unable to determine a $B$- or $V$-band magnitude, the optical flux would be lower therefore the ratio will be higher. The same applies to the brightest characterised counterparts, for although they have the lowest ratios, they are likely to be collections of less luminous objects which will all individually have higher ratios.

\section{Discussion}
\label{sec:disc}

The high luminosity of ULX-7 places it firmly into the category of ULXs, albeit at a luminosity that is not particularly remarkable within that class of sources. What makes ULX-7 remarkable is its unusual spectral and timing properties compared with the majority of ULXs. According to the classification of super-Eddington accretion regimes by \citet{sutton13}, sources in the hard ultraluminous regime have very low levels of variability if it is present at all, with high variability only featuring in sources in the soft ultraluminous regime. \citet{middleton15} suggests that the observed spectrum and variability of sources in ultraluminous accretion states depend on the inclination and accretion rate of the source, with the main driver of these differences being a radiatively-driven, massive and inhomogeneous wind, that imprints the variability on the hard component of the spectrum if it rises into the line-of-sight.

The energy spectrum of ULX-7 could be argued to be consistent with a hard ultraluminous accretion regime, with the characteristic two-component shape expected in the spectrum smeared out by insufficient data quality, except for a hint of a soft excess from {\it XMM-Newton} observation X3 and a putative high energy turnover in the {\it NuSTAR} data. However, were the source to truly be in this regime, we would not expect the high levels of variability that we observe across all observations. This is furthermore unusual given that variability is high at all observed energies, unlike the observed higher variability above 1\,keV in the soft ultraluminous regime. Additionally, the observed luminosity of ULX-7 varies by over an order of magnitude, but we see no evidence of the accretion properties changing. Therefore a soft, clumpy wind is unlikely to be the cause of this variability.

It is possible that our understanding of the hard ultraluminous regime of ULXs is as yet incomplete, and that ULX-7 is an unusually variable specimen of this accretion mode. However, as the data quality is insufficient to prefer a more complex spectral model over a power-law, we have examined other interpretations for the nature of this object that have power-law-like spectra, in particular considering the scenarios of a background AGN, a super-Eddington neutron star and an IMBH.

\subsection{Background AGN}
\label{sec:bgagn}

An occasional occurrence in studies of ULXs is the discovery that the object is a background AGN (e.g. \citealt{dadina13, sutton15}), rather than located within the galaxy it appears coincident with. The hard power-law spectrum that we observe is not inconsistent with ULX-7 being an AGN (e.g. \citealt{reeves00, mateos10}), so we need to look to other source properties to confirm its location.

Examining the possible optical counterparts suggests that this source is likely located within the galaxy. The sources that we are able to characterise are consistent either with OB supergiant type stars, or with clusters of cooler stars. While background quasars at intermediate redshifts could be consistent with the $B-V$ colour and magnitude, we find that for the OB-type stars the X-ray/optical ratios are high, with $F_{\rm X}/F_{\rm opt}>10$ in all cases. We would expect the majority of AGNs to have optical/X-ray flux ratios between 0.1 and 10 (e.g. \citealt{krautter1999, hornschemeier01}), so it is unlikely that any of these potential optical counterparts are background AGNs. We are only able to characterise the very brightest potential optical counterparts of ULX-7, given the limitations of the data, but since the X-ray/optical relationship would be even higher for the fainter objects we cannot characterise, these are even less likely to be a background AGN. 

The X-ray/optical relation alone comes with the caveat that we only have one epoch of $HST$ data and assume the same optical flux for the highest X-ray flux observation. If the optical emission also varies over time, this conclusion does not necessarily hold. It is also possible that a very highly obscured QSO would have an extreme X-ray/optical flux ratio. However, the proposal that ULX-7 is not a background AGN is also supported by the X-ray timing properties of the source, since there is a high amount of variability on timescales of $\sim100$\,s. This is shorter than expected for an AGN, for which noise tends to extend only down to timescales of tens of minutes to hours (e.g. \citealt{gonzalez12}). Additionally, the low-frequency break feature we see is also not often seen in AGNs -- one exception being Ark~564, which has a break at $7.5\times10^{-7}$\,Hz in the 2--8.8\,keV band \citep{mchardy07}, a much lower frequency than the one we see for ULX-7.

\subsection{Neutron Star}
\label{sec:ns}

Given the recent discovery that M82 X-2 is in fact a highly super-Eddington pulsar \citep{bachetti14}, another possible interpretation for ULX-7's unusual behaviour may be that it is a neutron star rather than a BH. To this end, we searched for coherent pulsations within the {\it XMM-Newton} data, but found no strong evidence for any between 6.85~Hz and 0.1~Hz with significance comparable to other studies (that is, with $H>23$). With that said, the absence of pulsations in an $H$-test do not necessarily mean that there is no stellar surface -- using a similar method, \citet{doroshenko15} were unable to detect pulsations from M82 X-2 in the {\it XMM-Newton} data for the source. It could instead mean that that either the pulsation amplitude was too low to be detected, or the spin-down rate and/or orbital modulation of the signal is significant enough to require an accelerated epoch folding search to detect pulsations.

The neutron star equivalent to super-Eddington accreting BHs are Z-sources, the most luminous neutron stars, accreting close to or above their Eddington limit \citep{hasinger89}. They can also exhibit high amounts of variability, although at very low frequencies their power spectra exhibit a steep power-law shape with $\alpha\sim$1--2, inconsistent with the power spectrum of ULX-7, which exhibits a low break and a flatter slope. A comparison with the very luminous extragalactic Z-source LMC X-2 further reveals that its energy spectrum is harder than that of ULX-7 as well \citep{barnard15}, so the properties of ULX-7 appear to be inconsistent with what we would expect from Z-sources.

However, the recently-reported spectral properties of M82 X-2 \citep{brightman15} indicate that it is possible for a super-Eddington neutron star ULX to show similar properties to ULX-7. As well as long-term flux variations over two orders of magnitude, examination of the pulsed spectrum in {\it NuSTAR} shows that the spectrum of M82 X-2 has a high energy turnover at $14^{+5}_{-3}$\,keV. Further observations using {\it NuSTAR} would help to confirm whether ULX-7 exhibits a similar spectral shape at high energies.

%Chandra results: power law slope of $\Gamma=1.33\pm0.15$ when fitted with a power-law and a black body disc with $T_{\rm in} = 3.24$\,keV

\subsection{Intermediate Mass Black Hole}
\label{sec:imbh}

Another possible interpretation is that ULX-7 is instead a BH accreting in a hard state analogous to lower luminosity BHBs. This would imply an unusually high BH mass due to its high luminosity, despite a low assumed accretion rate, and would manifest a hard power-law shaped spectrum with high variability across all energy bands (e.g. \citealt{grinberg14}) like we see in ULX-7. The irradiated disc of an IMBH would also be consistent with most of the possible optical counterparts we detect \citep{madhusudhan08}.

This interpretation is supported by the presence of a break in the power spectrum from a spectral index of $\alpha\sim0$ to $\alpha\sim-1$, a feature that we would expect from the low-frequency break in the power spectrum of a source in the hard state, which can be modelled by two Lorentzians or, more simply, a doubly-broken power-law (e.g. \citealt{done05}), whose high-frequency break scales with the BH mass. While we see no evidence of a high-frequency break, we can take the lower limit of such a break to be the white noise level of the {\it XMM-Newton} power spectrum, at $\nu_b = 9\times10^{-3}$\,Hz. We can use the relationship between high-frequency break and BH mass found to apply to BHs of all size scales by \citet{mchardy06}, with the offset for BHs in the hard state from \citet{kording07}, to calculate an upper limit on the BH mass using the following equation: $\log\nu_b = 0.98\log \dot{M}-2.1\log M_{\rm BH}-15.38$. We calculate $\dot{M}$ from $L_{\rm bol}/\eta c^2$, assuming an accretion efficiency of $\eta=0.1$ for the highest-flux observation, and obtain $L_{\rm bol}$ by applying a bolometric correction of 5 to the 2--10\,keV luminosity of that observation \citep{kording06}. In this way, we find an upper limit of $M_{\rm BH} < 1.6\times10^3$\,M$_{\odot}$, which means that ULX-7 is consistent with being an IMBH.

We would expect an IMBH accreting in the hard state to also exhibit radio emission from a jet. Since there is no radio detection of ULX-7, we can use the calculated flux density upper limit of 87\,$\mu$Jy\,beam$^{-­1}$ to establish an upper limit on the BH mass independent of that calculated from the timing analysis, using the fundamental plane in BH mass, radio and X-ray luminosity which has been found to apply to BHBs and AGNs as well as intermediate sources (e.g. \citealt{mezcua15}). We use the fundamental plane equation described in \citet{gultekin09}, which has been calibrated for low mass AGNs in the range $10^5-10^7$\,M$_{\odot}$ \citep{gultekin14}, and assume a flat radio spectral index of $\alpha=0.15$ to find $L_{5\rm\,GHz}$. We calculate a mass upper limit of $M_{\rm BH} < 3.5\times10^4$\,M$_{\odot}$, which also allows for an IMBH interpretation.

It is also possible to place a lower limit on the BH mass of an IMBH by taking the maximum observed flux of $8.6\times10^{-13}$\,erg\,cm$^{-2}$\,s$^{-1}$ and assuming a maximum accretion rate for the low/hard state, given that ULX-7 is a persistent source. The maximum luminosity of a low/hard state tends to be $\sim$2\% of Eddington\footnote{It is possible for transient BHs in the hard state to reach up to 100\% of Eddington before a transition to the soft thermal state \citep{dunn10}, however given that we see no evidence of state transitions for ULX-7, we calculate the lower limit using a more typical value for persistent hard states.}, with the highest Eddington ratios observed being $\sim$5\% \citep{maccarone03}. Therefore we use an Eddington ratio of 5\% to place a lower limit on the black hole mass of $M_{\rm BH} > 1.0\times10^3$\,M$_{\odot}$, which is consistent with our previously calculated upper limits and places the source firmly within the IMBH regime.

We can compare our results for ULX-7 with HLX-1, currently the best candidate for an IMBH due to its extreme luminosity and evidence of state transitions. When first discovered, it had a spectrum consistent with an absorbed power-law \citep{farrell09}, albeit a softer one than we see in ULX-7. Further studies have revealed it to have a very high dynamic range, as we see for ULX-7, although its spectrum changes shape and it appears to demonstrate state transitions \citep{godet09} whereas ULX-7 appears to remain in a single state. In its third {\it XMM-Newton} observation, HLX-1 appeared to enter a hard state, with a lower luminosity and a spectral index of $\Gamma=1.6\pm0.4$ when compensating for the host galaxy's contribution to the soft emission and fitted alongside an accretion disc \citep{servillat11}. No significant intrinsic variability was detected, although since the power was not well constrained, a high fractional variability was not ruled out. Additionally, there have also been radio detections of HLX-1 while in this state \citep{cseh15}, making it analogous to the hard state seen in stellar-mass BHBs. 

From this, we can conclude that it is reasonable to suggest that ULX-7 could also be an IMBH in a hard state, its large mass being the cause of its high luminosities, although unlike HLX-1, it does not appear to undergo state transitions as it changes luminosity.

The association of ULX-7 with a young stellar population implies that it is a short-lived source if it was formed there \citep{roberts07}, and in this respect it bears similarity to the wider ULX population which is found predominantly in star-forming regions. This would be a point in favour of a more standard stellar remnant ULX interpretation. However, there are possible formation scenarios for an IMBH in a young stellar environment. For example, an IMBH could have formed through runaway mergers within a dense stellar cluster and subsequently been ejected, or the cluster dissipated into the disc of the galaxy, leaving the IMBH accreting within a dense molecular cloud (e.g. \citealt{miller02}) or retaining a young stellar population around itself (e.g. \citealt{farrell12}). This is an unlikely formation scenario for the ULX population as a whole \citep{king04}, but still a possibility for an individual object, as a very rare occurence. 

Another possibility is a minor merger of a dwarf galaxy with the main galaxy, a mechanism that has been suggested for HLX-1 \citep{farrell12, mapelli12} and NGC 2276-3c \citep{mezcua15}. A recent minor merger could be identified by evidence of disruption in the spiral arm around the source and increased levels of star formation, however these are seen in the northern spiral of M51 anyway due to M51a's interaction with M51b. Therefore any evidence for a minor merger that could have formed ULX-7 would likely be eclipsed by the disruption of the current interaction.

\section{Conclusions}
\label{sec:conc}

We have undertaken a case study of M51 ULX-7, a source with moderate luminosity and very high variability for a ULX, and a consistently hard spectrum. This is in contrast to expected ULX variability behaviour, in which we might expect to see high variability in sources with soft spectra. We find that the source is generally well-fitted by a power-law with a spectral photon index that remains steady at $\Gamma\sim1.5$ while the source luminosity varies by over an order of magnitude over the course of 12 years. ULX-7 also demonstrates very high fractional variability between 0.3 and 10.0\,keV, with a broken power-law shape to its power spectrum analogous to the low-frequency break in the power spectrum of an X-ray binary accreting in the hard state. We find solid evidence for a positive linear rms-flux relation, making ULX-7 the third ULX for which this feature is confirmed. We find no evidence of coherent pulsations, however. 

Taken together, these properties are unusual for a ULX, and are suggestive of an alternative explanation to the broadened disc or ultraluminous regimes that describe the majority of ULXs for which we have reasonable data \citep{gladstone09, sutton13}. By examining the possible optical counterparts in {\it HST} data, we consider it unlikely that this source is a background AGN. The lack of pulsations and dissimilarity to Z-source properties imply that it is not a neutron star either, although it may possibly bear similarities to the properties of the neutron star ULX M82 X-2.

Our results are consistent with ULX-7 being an IMBH accreting in the hard state. Using the absence of a high-frequency break and a radio detection, we can calculate upper limits on the BH mass of $M_{\rm BH} < 1.55\times10^3$\,M$_{\odot}$ and $M_{\rm BH} < 3.5\times10^4$\,M$_{\odot}$ respectively, and by taking the maximum accretion rate to be 5\% of the Eddington limit, we can calculate a lower mass limit of $M_{\rm BH} > 1.0\times10^3$\,M$_{\odot}$. All of these limits show the source to be consistent with an IMBH interpretation if we assume it is accreting in the hard state. There remains weak evidence of a possible high energy turnover in the spectrum when considering the {\it NuSTAR} data on this source, which would imply that this source may instead be exhibiting some permutation of the ultraluminous state after all, but simultaneous deep observations with {\it XMM-Newton} and {\it NuSTAR} will be required to confirm or rule out the existence of these features.

\section*{Acknowledgements}

We gratefully acknowledge support from the Science and Technology Facilities Council (HE through grant ST/K501979/1 and TR as part of consolidated grant ST/L00075X/1) and from NASA (MM through Chandra Grant G05-16099X). AS is supported by an appointment to the NASA Postdoctoral Program at Marshall Space Flight Center, administered by Oak Ridge Associated Universities through a contract with NASA.

The scientific results reported in this paper are based on data obtained from the {\it Chandra} Data Archive, and on archival observations obtained with {\it XMM-Newton}, an ESA science mission with instruments and contributions directly funded by ESA Member States and NASA. This research has also made use of data obtained with \textit{NuSTAR}, a project led by Caltech, funded by NASA and managed by NASA/JPL, and has utilized the \textsc{NuSTARDAS} software package, jointly developed by the ASDC (Italy) and Caltech (USA). Further results are based on observations made with the NASA/ESA Hubble Space Telescope, and obtained from the Hubble Legacy Archive, which is a collaboration between the Space Telescope Science Institute (STScI/NASA), the Space Telescope European Coordinating Facility (ST-ECF/ESA) and the Canadian Astronomy Data Centre (CADC/NRC/CSA); and the National Radio Astronomy Observatory, which is a facility of the National Science Foundation operated under cooperative agreement by Associated Universities, Inc.

%One copy of each paper resulting from data obtained from the HLA should be sent to the STScI. 

\bibliography{m51apaper}
\bibliographystyle{mn2e}
\bsp

\label{lastpage}

\end{document}